\documentclass[11pt]{article}

\usepackage{url,amsmath,amscd,amsfonts,graphicx,psfrag,marginnote,hyperref,amssymb,bm,mathtools}
\usepackage{amsthm}
\usepackage[all]{xy}
\usepackage{color}
\usepackage{fancyhdr}
\usepackage{wrapfig}

\newtheoremstyle{stuff}{\topsep}{\topsep}%
     {\itshape}
     {}
     {\bfseries}
     {}
     {.5em}
     {\thmnote{#3}}

\theoremstyle{plain}
\newtheorem{thm}{Theorem}

\newtheorem{lem}[thm]{Lemma}
\newtheorem{prop}[thm]{Proposition}

\theoremstyle{remark}
\newtheorem*{rem}{Remark}

\theoremstyle{definition}
\newtheorem{defn}{Definition}

\newtheorem{example}{Example}

\theoremstyle{stuff}

\numberwithin{equation}{section}
 
\usepackage[scale=0.72]{geometry}

\usepackage{sectsty}
\sectionfont{ \bf \Large}
\subsectionfont{\bf \large}
\subsubsectionfont{\it \large}

\usepackage{setspace}

\newcommand{\bea}{\begin{eqnarray}}
\newcommand{\eea}{\end{eqnarray}}
\newcommand{\be}{\begin{equation}}
\newcommand{\ee}{\end{equation}}

\begin{document}

\pagestyle{empty}

\begin{flushright}
\begin{tabular}{l}
 IPHT--T12/114
\end{tabular}
\end{flushright}

\vspace*{1.5cm}

{\noindent \Large \bf\fontfamily{pag}\selectfont Think globally, compute locally}\\

\vspace*{0.15cm}
\noindent \rule{\linewidth}{0.5mm}

\vspace*{1cm}

{\noindent \fontfamily{pag}\selectfont  Vincent Bouchard$^\dagger$ and Bertrand Eynard$^\circ$}\\[2em]
{\small \it  \indent $^\dagger$ Department of Mathematical and Statistical Sciences\\
\indent University of Alberta\\
\indent 632 CAB, Edmonton, Alberta T6G 2G1\\
\indent Canada}\\
\indent \url{vincent@math.ualberta.ca}\\[1em]
{\small \it
 \indent $^\circ$ Institut de Physique Th\'eorique\\
\indent CEA Saclay}\\
\indent \url{bertrand.eynard@cea.fr}\\

\vspace*{1cm}

\hspace*{1cm}
\parbox{11.5cm}{{\sc Abstract:} We introduce a new formulation of the so-called topological recursion, that is defined globally on a compact Riemann surface. We prove that it is equivalent to the generalized recursion for spectral curves with arbitrary ramification. Using this global formulation, we also prove that the correlation functions constructed from the recursion for curves with arbitrary ramification can be obtained as suitable limits of correlation functions for curves with only simple ramification. It then follows that they both satisfy the properties that were originally proved only for curves with simple ramification.}
\vspace*{1cm}


\vspace*{0.5cm}


\pagebreak

\pagestyle{fancy}
\pagenumbering{arabic}

\tableofcontents 


%

\section{Introduction}

The topological recursion proposed in \cite{Eynard:2007} appears in many counting problems in enumerative geometry. It has found applications in areas as diverse as Gromov-Witten theory, topological string theory, matrix models, mirror symmetry, Hurwitz theory, knot theory, Seiberg-Witten theory, and many more \cite{BE,BEMS,BCMS,BKMP1,BKMP2,BM,BEM,Chen:2009,DFM,DMSS,Eynard:2011,Eynard:2011ii,EMS:2009,Eynard:2007,Eynard:2008,EO:2012,GS,HK,Marino:2008,NS:2010,NS:2011,Zhou:2009,Zhou:2009ii,Zhu:2011}.

The recursion relies on the geometry of a compact Riemann surface $S$ with two meromorphic functions $x$ and $y$ generating the function field $K(S) = \mathbb{C}(x,y)$. To formulate the recursion, we realize $S$ as a branched covering $x: S \to \mathbb{C} \mathbb{P}^1$. In the original formulation of \cite{Eynard:2007,Eynard:2008}, it was required that $x$ be ``generic'', in the sense that it only has simple ramification points. While it is well known that any branched cover with higher ramification can be obtained as a suitable limit of a branched cover with only simple ramification, until recently it was not clear how to apply the recursion to branched covers with higher ramification. A generalization of the recursion for spectral curves with higher ramification is desirable, because in some counting problems the spectral curve does indeed have higher ramification.

Such a generalization was proposed in \cite{BHLMR}, based on the work of Prats-Ferrer \cite{Prats:2010} in matrix models. The generalized recursion was studied in a number of examples, and it was shown that it restores $x-y$ invariance for spectral curves where $x$ has only simple ramification but $y$ has higher ramification. It was also computed, in examples, that many of the properties of the correlation functions proved in \cite{Eynard:2007, Eynard:2008} are also satisfied by the correlation functions constructed from the generalized recursion. However, the evidence for the generalized recursion was mostly computational; it was not proved that the proposed generalization was the appropriate one. This is what we address in this paper.

More precisely, an obvious question is the following. Any branched cover with higher ramification can be obtained as a suitable limit of a branched cover with only simple ramification, where in the limiting process ramification points collide to produce higher order ramification points. Suppose that we construct the correlation functions from the original recursion applied to the curve with only simple ramification. Is the limit of these correlation functions equal to the correlation functions constructed using the generalized recursion applied to the limiting curve with higher order ramification? If the generalization is the correct one, then the answer should be affirmative.

In this paper we prove that indeed, the generalized recursion is a limit, in the sense above, of the original recursion of \cite{Eynard:2007,Eynard:2008}; this follows from our main result, Theorem \ref{t:main}. In order to prove such a statement, we develop a new formulation of the recursion, which we call ``global'', for the following reason.

The recursion involves contour integrals around ramification points of the branched cover $x:S \to \mathbb{C} \mathbb{P}^1$. In both the original and generalized recursions, the integrand is only defined locally in a small disk around the ramification points. While this is perfectly sufficient to perform the contour integrations, it makes studying global properties of the recursion rather difficult. For instance, it is hard to study limiting process where ramification points collide, since for this we need an integrand defined away from a given ramification point.

To perform global studies of the recursion, we introduce a new recursion, which also involves contour integration around ramification points, but where the integrand is now a globally defined meromorphic differential on $S$. We show that this ``global'' recursion is in fact precisely equivalent to the original and generalized recursions (which we call ``local''), which is our main result (Theorem \ref{t:main}). From this theorem, and the global nature of the recursion, it follows that the limits of the correlation functions when ramification points collide are indeed equal to the correlation functions constructed from the limiting curve.

Using this main theorem, we also prove that the correlation functions constructed for spectral curves with arbitrary ramification satisfy the same properties as the original correlation functions constructed in \cite{Eynard:2007,Eynard:2008}.

It would be interesting to investigate further whether this new, global, recursion provides insights on other aspects of the recursion, such as $x-y$ invariance of the free energies, which was first investigated in \cite{EOxy} (see also \cite{BS:2011,BHLMR} for a discussion of $x-y$ invariance). We hope to report on that in the near future.

\subsection*{Outline}

In section 2 we introduce the geometry, and construct the generalized recursion of \cite{BHLMR} and its special case for simple ramification introduced originally in \cite{Eynard:2007,Eynard:2008}. We also prove some properties of the recursion that will be useful later on. In section 3, we construct a global recursion, whose integrand is a globally defined meromorphic function. We then prove our main result, Theorem \ref{t:main}, which says that the global recursion is in fact precisely equal to the local formulation in \cite{BHLMR}. In section 4, we show that Theorem \ref{t:main} implies that the correlation functions for curves with arbitrary ramification satisfy the same properties that were proved in \cite{Eynard:2007,Eynard:2008}. In section 5 we conclude and discuss some future avenue of research. Finally, in the Appendix we prove one of the properties discussed in section 4 explicitly, partly to highlight the fact that it is highly non-trivial that the correlation functions for arbitrary ramification satisfy these properties.

\subsection*{Acknowledgments}

We would like to thank M.~Mulase for very enjoyable and useful discussions, and the referee for interesting comments. V.B. would also like to thank J.~Hutchinson, P.~Loliencar, C.~Marks, M.~Meiers and M.~Rupert for interesting discussions. We would like to give special thanks to the American Institute of Mathematics for funding a very productive SQuaREs meeting, where this work was initiated. The research of V.B. is supported by an NSERC Discovery grant.

\section{The local topological recursion}

\label{s:EO}

In this section we review the construction of the original topological recursion formulated in \cite{Eynard:2007, Eynard:2008}, and its generalization proposed in \cite{BHLMR}. We will use the notation put forward by Prats Ferrer in \cite{Prats:2010} and used in the generalized context in \cite{BHLMR}. We will call the generalized recursion of \cite{BHLMR} the ``local'' topological recursion, to contrast with the formulation of the recursion that will be proposed in the next section.

In the last subsection we will also prove a number of properties of the local recursion that will be useful later on.

\subsection{The geometry}

\subsubsection{Spectral curves}

The main object of study will be called a \emph{spectral curve}.

\begin{defn}\label{d:spectral}
A \emph{spectral curve} $C$ is a triple $C=(S,x,y)$ where $S$ is a Torelli marked\footnote{A Torelli marked compact Riemann surface $S$ is a compact Riemann surface $S$ with a choice of symplectic basis $(A_1, \ldots, A_g, B_1, \ldots B_g)$ for $H_1(S, \mathbb{Z})$.} compact Riemann surface, and $x$ and $y$ are meromorphic functions on $S$ that generate the function field of $S$, that is, $K(S) = \mathbb{C}(x,y)$. We require that $x$ and $y$ separate tangents, in the sense that for any $p \in S$, either $\mathrm{d} x(p) \neq 0$ or $\mathrm{d} y(p) \neq 0$. We let $d$ be the degree of $x$.
\end{defn}

Note that any two such functions $x$ and $y$, where $x$ is of degree $d$, satisfy identically an equation (which can always be taken to be irreducible) of the form
\begin{equation}
y^d + a_{d-1}(x) y^{d-1} + \ldots + a_0(x) = 0,
\end{equation}
where $a_i(x) \in \mathbb{C}(x)$, $i=0, \ldots, d-1$. If one takes the point of view that $x$ and $y$ are coordinates on $\mathbb{C}^2$ instead of meromorphic functions on $S$, then this equation defines an affine curve in $\mathbb{C}^2$ (after turning it into a polynomial equation), whose normalization is the compact Riemann surface $S$.

\subsubsection{Branched cover}

In the following we will study the spectral curves through the degree $d$ branched covering $x: S \to \mathbb{C} \mathbb{P}^1$ given by the meromorphic function $x$. Given a point $p \in S$, then there exists local coordinates near $p \in S$ and $x(p) \in \mathbb{C}_\infty$ such that $x$ can be written as $z \mapsto z^m$ for a positive integer $m \leq d_x$. $m$ is uniquely defined, and is called the \emph{multiplicity} of $x$ at $p$; it is denoted by $\text{mult}_p(x)=m$. All but a finite number of points have $\text{mult}_p(x)=1$. The points with $\text{mult}_p(x) \geq 2$ are the \emph{ramification points} of $x$. Their images in $\mathbb{C} \mathbb{P}^1$ under the $x$-map are the \emph{branch points} of $x$.

The multiplicity $\text{mult}_p(x)$ of $x$ at $p$ can also be understood in terms of the meromorphic function $x(p)$ for $p \in S$. If $p$ is not a pole of $x$, then $\text{mult}_p(x) = \text{ord}_p(x-x(p))$. Equivalently, in terms of the differential $\mathrm{d} x$, we obtain $\text{mult}_p(x) = \text{ord}_p(\mathrm{d} x) + 1$. If $p$ is a pole of $x$, then $\text{mult}_p(x) = - \text{ord}_p(x) = - \text{ord}_p( \mathrm{d} x)-1$. Therefore, ramification points of $x$ correspond to zeroes of $\mathrm{d} x$ and poles of $x$ of order $2$ or more.  

In the following, we will only focus on ramifications points of $x$ corresponding to zeros of $\mathrm{d} x$.\footnote{We could include the ramification points corresponding to poles of $x$ in the discussion, but they would not contribute to the recursion, hence are irrelevant for our purposes.} Let $R = \{ a_1, \ldots, a_n\} \subset S$ be the set of zeroes of $\mathrm{d} x$, with corresponding multiplicities $r_i = \text{mult}_{a_i}(x)$, for $i=1,\ldots,n$, and $B = \{b_1, \ldots, b_n \} := \{x(a_1), \ldots, x(a_n)\} \subset \mathbb{C} \subset \mathbb{C} \mathbb{P}^1$ be the set of associated branch points. \emph{A priori}, the $b_i$ need not be distinct, but when we formulate the ``global'' recursion in the next section we will need to require that they are.

\begin{rem}
Note that in the original case studied in \cite{Eynard:2007,Eynard:2008},  it was assumed that all zeroes of $\mathrm{d} x$ were simple, that is, $r_i=2$ for $i=1,\ldots,n$. The general case for $r_i$ not necessarily equal to $2$ was studied in \cite{BHLMR}.
\end{rem}

Let $U_i$ be a small disk centered at $a_i$ that does not include other $a_j \neq a_i$, and $V_i$ be a small disk centered at $b_i$. The restriction $x: U_i \to V_i$ is a degree $r_i$ Galois cover, fully ramified at $a_i$. Since, locally on $U_i$, $x$ is a Galois cover, there is a non-trivial deck transformation group on $U_i$, which is just the cyclic group $\mathbb{Z} /{ r_i} \mathbb{Z}$. We denote by $\theta_i: U_i \to U_i$ its generator, which satisfies $x \circ \theta_i = x$ by definition.  In the following we will use the notation $\mathrm{d}_i(p) = \{ \theta_i(p), \theta_i^2(p), \ldots, \theta_i^{r_i-1}(p) \}$ to denote the non-trivial images of $p \in U_i$ under the action of the cyclic deck transformation group. 

Note that deck transformations can be pulled back to meromorphic functions on $U_i$. Let $f(p)$ be a meromorphic function on $U_i$, and $\theta_i :U_i \to U_i$ be a deck transformation. Then $\theta_i^* f(p) = (f \circ \theta_i)(p)$ is a well defined meromorphic function on $U_i$. We will denote the pullbacked function $\theta_i^* f(p)$ by $f(\theta_i(p))$.

\begin{rem}
It is very important to realize that this analysis can only be done locally on $U_i$ near a ramification point $a_i$. In most cases, the branched covering $x: S \to \mathbb{C} \mathbb{P}^1$ will not globally be a Galois cover, and in fact its deck transformation group may be trivial. But since in the recursion we will consider contour integrals around the $a_i$, a local analysis is sufficient to formulate the recursion. However, it will be desirable to write the recursion in a global context, which we will do in the next section. For this however we will not be able to use the local deck transformations.
\end{rem}

\subsubsection{Examples of spectral curves}

Let us give three simple examples of genus $0$ spectral curves.

\begin{example}
Let $S = \mathbb{C} \mathbb{P}^1$, and $x$ and $y$ be the meromorphic functions defined in terms of an affine coordinate $t \in \mathbb{C} \mathbb{P}^1$ by:
\begin{equation}
x(t) = t^5, \qquad y(t) = t.
\end{equation}
The corresponding equation is
\begin{equation}
y^5 - x = 0,
\end{equation}
which defines a smooth affine curve whose normalization is $\mathbb{C} \mathbb{P}^1$. 

The degree $5$ branched covering $x: S \to \mathbb{C} \mathbb{P}^1$ is ramified at $t=0$ (and at $\infty$). Let $a = \{t=0\} \in S$, which has multiplicity $r=5$. On $U_a$, the deck transformation group is $\mathbb{Z} / 5 \mathbb{Z}$. In the local coordinate $t$, the generator of the deck transformation group is $t \mapsto \omega t$, where $\omega$ is a primitive $5$'th root of unity. In fact, in this example the branched covering $x: S \to \mathbb{C} \mathbb{P}^1$ is globally a Galois cover, and the deck transformations are globally defined on $S$.

Note that if we had instead chosen
\begin{equation}
x(t) = t^5, \qquad y(t) = t^2,
\end{equation}
then it would not be a spectral curve according to our definition, since while $x$ and $y$ generate $K(\mathbb{C} \mathbb{P}^1)$, they have a common zero at $t=0$.
\end{example}

\begin{example}\label{ex:2}
Let $S = \mathbb{C} \mathbb{P}^1$, and $x$ and $y$ be the meromorphic functions defined in terms of an affine coordinate $t \in \mathbb{C} \mathbb{P}^1$ by:
\begin{equation}
x(t) = t + \frac{1}{t}, \qquad y(t) = t^2.
\end{equation}
The corresponding equation is
\begin{equation}
y^2 + y (2-x^2) + 1   = 0,
\end{equation}
which defines an affine curve that has an ordinary double point at $(x,y) = (0,-1)$, and whose normalization is $\mathbb{C}\mathbb{P}^1$. While the affine curve is singular, this is a perfectly well defined spectral curve, since the zeroes of $\mathrm{d} x$ and $\mathrm{d} y$ do not coincide.

The degree $2$ branched covering $x: S \to \mathbb{C} \mathbb{P}^1$ is ramified at $t=\pm 1$, which we denote by $a_{\pm} \in S$. On both $U_{a_{\pm}}$ the deck transformation group is $\mathbb{Z}/2 \mathbb{Z}$ generated by $t \mapsto 1/t$. Again, in this case the deck transformations are globally defined on $S$.
\end{example}

\begin{example}\label{ex:bad}
Let $S = \mathbb{C} \mathbb{P}^1$, and $x$ and $y$ be the meromorphic functions defined in terms of an affine coordinate $t \in \mathbb{C} \mathbb{P}^1$ by:
\begin{equation}
x(t) =\frac{t^3}{3} - t, \qquad y(t) = (t+1)(t-2).
\end{equation}
The corresponding equation is
\begin{equation}
y^3 + 3 y( 3x-2) - (3x-2)^2 = 0.
\end{equation}
It defines an affine curve which has an ordinary double point at $(x,y) = \left( \frac{2}{3}, 0 \right)$.

The degree $3$ branched covering $x: S \to \mathbb{C} \mathbb{P}^1$ has two simple ramification points at $t = \pm 1$, which we denote by $a_{\pm} \in S$ (it is also ramified at $\infty$). On both $U_{a_{\pm}}$ the deck transformation group is $\mathbb{Z} / 2 \mathbb{Z}$; here however $x$ is not globally a Galois cover, and in fact the global deck transformation group is trivial. If we use the coordinate $t$ above, restricting to $U_{a_+}$ the local deck transformation has a series expansion near $t=1$ of the form
\begin{equation}
t \mapsto 1 - (t-1) + \mathcal{O}(t-1)^2,
\end{equation}
while on $U_{a-}$ near $t=-1$ the local deck transformation has a series expansion
\begin{equation}
t \mapsto  -1 - (t+1) + \mathcal{O}(t+1)^2.
\end{equation}

While this curve is a perfectly well defined spectral curve, it is peculiar in the following sense. It turns out that one of the ramification point of $x$, namely $a_-$ at $t=-1$, is in the preimage of the double point singularity of the affine curve under its normalization. Another way of putting it is as follows. Let $b_- = x(-1) \in \mathbb{C} \mathbb{P}^1$ be the branch point corresponding to $a_-$. Then there exists a $p \neq a_- \in S$ in the preimage of $x^{-1}(b_-)$ such that $(x(a_-), y(a_-)) = (x(p), y(p))$ -- in our case, $p$ is given by the point $t=2$. What that means is that the normalization of the affine curve is not a bijection at $(x(a_-), y(a_-))$, that is, the affine curve is singular at $(x(a_-), y(a_-))$.

This peculiarity is not problematic \emph{a priori}, but when we introduce a ``global'' version of the recursion in the next section we will need to make the further assumption that such spectral curves do not occur.

\end{example}

\subsubsection{Meromorphic differentials}

Given a spectral curve $(S,x,y)$, the  type of objects that we will be constructing are \emph{symmetric meromorphic differentials}.

\begin{defn}[\cite{MS:2012}]
Let $D$ be a divisor of $S$, and denote the canonical bundle on $S$ by $K_S$. A \emph{symmetric meromorphic differential} $W_n(p_1, \ldots, p_n)$ of degree $n$ on $S$ with poles along $D$ is an element of the symmetric tensor product
\begin{equation}
W_n(p_1, \ldots, p_n) \in \text{Sym}^n H^0(S, K_S(* D) ).
\end{equation}
\end{defn}

We will also need a particular bilinear differential on the Cartesian product $S \times S$, which is a classical object in the theory of Riemann surfaces.

\begin{defn}
Let $\triangle \subset S \times S$ be the diagonal, and $\pi_1, \pi_2 : S \times S \to S$ be the projections on both factors. The \emph{canonical bilinear differential of the second kind}
\begin{equation}
W_{2}^0(p_1,p_2) \in H^0(S \times S, \pi_1^* K_S \otimes \pi_2^* K_S \otimes \mathcal{O}(2 \triangle))
\end{equation} 
is the unique bilinear differential on $S \times S$ defined by the conditions:
\begin{itemize}
\item It is symmetric, $W_{2}^0(p_1, p_2) = W_{2}^0(p_2,p_1)$;
\item It has its only pole, which is double, along the diagonal $p_1 = p_2$, with no residue; in local coordinates $z_1 := z(p_1)$ and $z_2 := z(p_2)$, its expansion in this neighborhood has the form
\begin{equation}
W_{2}^0(p_1, p_2) = \left(\frac{1}{(z_1-z_2)^2} + \text{regular} \right) \mathrm{d} z_1 \mathrm{d} z_2.
\end{equation}
\item It is normalized about a symplectic basis of $(A^I, B_J)$ cycles on $S$ such that
\begin{equation}
\oint_{A^I} W^0_2( \cdot, p) = 0.
\end{equation}
\end{itemize}
\end{defn}

\begin{rem}
Note that given a spectral curve $(S,x,y)$, the canonical bilinear differential $W^0_2(p_1,p_2)$ is uniquely defined. In fact, it only depends on $S$ and the choice of canonical basis of cycles, hence two spectral curves $(S,x,y)$ and $(S, \tilde{x}, \tilde{y})$ with the same Torelli marked compact Riemann surface $S$ have the same canonical bilinear differential.
\end{rem}

\subsection{The local recursion}

\subsubsection{Notation}
\label{s:notation}

Before we move on with the definition of the recursion, let us fix the notation that will be used in the remaining of this section.

Let $C=(S,x,y)$ be a spectral curve. We denote by $R = \{a_1, \ldots, a_m\}$ the set of zeroes of $\mathrm{d} x$, with corresponding multiplicities $r_i = \text{mult}_{a_i}(x)$. For each ramification point $a_i \in R$, let $\mathbf{d}_i(q)= \{\theta_i^{1}(q), \ldots, \theta_i^{r_i-1}(q) \}$ be the non-trivial images of $q \in U_{a_i}$ under the local deck transformation group. The notation $\mathbf{d}_i'(q) \subseteq \mathbf{d}_i(q)$ denotes a non-empty subset of $\mathbf{d}_i(q)$.

To denote points on the Riemann surface $S$, we will use the letters $p$ and $q$. For instance, $p_0 \in S$, or $q_2 \in S$. We will also use the set notation $\mathbf{p} = \{ p_1, \ldots, p_n \} \in S^n$ and $\mathbf{q}=\{ q_1,\ldots, q_k \} \in S^k$.

\subsubsection{The local recursion}

Now given a spectral curve $C$, we construct recursively an \emph{infinite tower of symmetric meromorphic differentials}. More precisley, let $\{ W_{n}^g \}$ be an infinite sequence of degree $n$ symmetric meromorphic differentials $W_{n}^g(p_1, \ldots, p_n)$ for all integers $g\geq 0$ and $n > 0$ subject to $2g - 2 + n > 0$, with poles only along $R$. Let $W_{2}^0(p_1, p_2)$ be the canonical bilinear differential of the second kind on $S \times S$. We will sometimes call these differentials \emph{correlation functions}.

Recall that a partition of a set $X$ is a collection of nonempty subsets of $X$ such that every element of $X$ is in exactly one of these subsets. To fix notation, a partition $\mu$ of $X$ is a collection of $s = \ell(\mu)$ subsets $\mu_i \subseteq X$, $i=1,\ldots,s$, such that $\cup_{i=1}^s \mu_i = X$ and $\mu_i \cap \mu_j = \emptyset$ for all $i \neq j$. We denote by $|\mu_i|$ the number of elements in the subset $\mu_i$; clearly $\sum_{i=1}^s |\mu_i| = |X|$ and $|\mu_i| \neq 0$ for all $i=1,\ldots,s$. The trivial partition is of course $\mu = \{ X \}$, which has length $s = \ell(\mu ) = 1$, and $\mu_1 = X$.

\begin{defn}\label{d:curly}
We define:
\begin{equation}\label{e:curly}
\mathcal{W}_{k,n}^g(\mathbf{q}; \mathbf{p}) = \sum_{\mu \in \text{Partitions}(\mathbf{q})} \sum'_{\substack{\sum_{i=1}^{\ell(\mu)} g_i = g + \ell(\mu) - k \\ \cup_{i=1}^{\ell(\mu)} J_i =  \mathbf{p}}} \left( \prod_{i=1}^{\ell(\mu)} W^{g_i}_{|\mu_i| + |J_i|} (\mu_i, J_i)  \right).
\end{equation}
The first summation is over partitions $\mu$ of the set $\mathbf{q}$. The second summation involves summing over all possible $\ell(\mu)$-tuples of non-negative integers $(g_1, \ldots, g_{\ell(\mu)})$, where $\ell(\mu)$ is the number of subsets in the partition $\mu$, satisfying $\sum_{i=1}^{\ell(\mu)} g_i = g + \ell(\mu) - k$. It also involves summing over all possible non-intersecting subsets $J_i \subseteq \mathbf{p}$, for $i=1, \ldots, \ell(\mu)$, such that $\cup_{i=1}^{\ell(\mu)} J_i =  \mathbf{p}$. Finally, the prime over the second summation symbol means that we exclude the cases with $(g_i, |\mu_i| + |J_i|) = (0,1)$ for some $i$.

Note that by definition, $\mathcal{W}_{k,n}^g(\mathbf{q}; \mathbf{p})$ is symmetric in all its variables.
\end{defn}

We now define the kernel that will enter into the recursion:
\begin{defn}\label{d:kernel}
The \emph{recursion kernel} $K_k(p_0;\mathbf{q})$ is defined by
\begin{equation}
K_k(p_0;\mathbf{q}) = - \frac{dS_{q_1,o}(p_0)}{\prod_{i=2}^{k} \omega(q_1, q_i)},
\end{equation}
where $dS_{q_1,o}(p_0)$ is the canonical normalized Abelian differential of the third kind on $S$: 
\begin{equation}
dS_{q_1,o}(p_0) = \int_o^{q_1} W^0_2(p_0, \cdot),
\end{equation}
with $o \in S$ an arbitrary base point (it has simple poles at $p_0=q_1$ and $p_0=o$ with respective residues $+1$ and $-1$), and $\omega(q_1,q_i)$ is defined in terms of the functions $x$ and $y$ by
\begin{equation}
\omega(q_1,q_i) = (y(q_1) - y(q_i)) \mathrm{d} x(q_1).
\end{equation}
Note that by definition, $K_k(p_0; \mathbf{q})$ is symmetric under permutations of $q_2, \ldots, q_k$ (but not under permutations involving $q_1$).
\end{defn}

Putting this together, we define the local recursion, which was termed \emph{generalized topological recursion} in \cite{BHLMR}:
\begin{defn}\label{d:recursion}
Let $C$ be a spectral curve. Using the notation defined in subsection \ref{s:notation}, we uniquely construct degree $n$ symmetric meromorphic differentials $W_n^g$ on $S$ with poles along $R$ through the \emph{local recursion}:
\begin{equation}\label{e:recursion}
W^g_{n+1}(p_0, \mathbf{p}) = \frac{1}{2 \pi i} \sum_{i=1}^m \oint_{\Gamma_i(q)} \left( \sum_{\mathbf{d}_i'(q)}  K_{|\mathbf{d}_i'|+1}(p_0;q,\mathbf{d}_i'(q)) \mathcal{W}^g_{|\mathbf{d}_i'|+1,n}(q, \mathbf{d}_i'(q); \mathbf{p} ) \right).
\end{equation}
The second summation is over all non-empty subsets of $\mathbf{d}_i(q)= \{\theta_i^{1}(q), \ldots, \theta_i^{r_i-1}(q) \}$ for each ramification points $a_i \in R$. The contour integrals are in $q$, over small circles $\Gamma_i(q)$ around each ramification point. In a local coordinate $z(q)$ on $U_i$, the contours are defined by $\Gamma_i(q) = \{ |z(q)-z(a_i)| = \epsilon_i \}$ for some small $\epsilon_i$.
\end{defn}

We also define:
\begin{defn}
The \emph{free energies} $F_g$, $g \geq 2$, are uniquely constructed from the one-forms $W_1^g(q)$ by:
\begin{equation}
F_g \coloneqq W^g_0 = \frac{1}{2g-2} \left( \frac{1}{2 \pi i} \sum_{i=1}^m \oint_{\Gamma_i(q) } \Phi(q) W_1^g(q) \right),
\end{equation}
where $\Phi(q)$ is a primitive of the one-form $y \mathrm{d} x$:
\begin{equation}
\Phi(q) = \int^q_o y(t) \mathrm{d} x (t),
\end{equation}
where $o \in S$ is an arbitrary base point. Note that $\Phi(q)$ may not be a meromorphic function on $S$, but it is meromorphic on the disks $U_i$ near the ramification points $a_i$.
\end{defn}

\subsubsection{The original recursion of \cite{Eynard:2007, Eynard:2008}}

The original recursion introduced in \cite{Eynard:2007, Eynard:2008} is a special case of the local recursion presented in the previous subsection, when all zeroes of  $\mathrm{d} x$ are simple. In this case, at each $a_i$ the local deck transformation group is $\mathbb{Z}/2 \mathbb{Z}$, and $p \in U_{a_i}$ has a unique non-trivial image $\theta_i(p)$ under the action of local deck transformations, since locally $x$ is a two-sheeted branched covering. Thus, in the definition of the recursion, \eqref{e:recursion}, the sum over subsets of $\mathbf{d}(q)$ collapses, and we get the formula
\begin{equation}\label{e:recursionEO}
W^g_{n+1}(p_0, \mathbf{p}) =  \frac{1}{2 \pi i} \sum_{i=1}^m \oint_{\Gamma_i(q)} \left(  K_{2}(p_0;q,\theta_i(q)) \mathcal{W}^g_{2,n}(q, \theta_i(q); \mathbf{p} ) \right).
\end{equation}
Recalling the definition of the kernel, Definition \ref{d:kernel}, and of the curly $\mathcal{W}$, Definition \ref{d:curly}, we can rewrite the recursion as
\begin{align}
W^g_{n+1}(p_0, \mathbf{p}) =  -  \frac{1}{2 \pi i} \sum_{i=1}^m \oint_{\Gamma_i(q)} \frac{dS_{q,o}(p_0)}{\omega(q,\theta_i(q))}& \Bigg( W^{g-1}_{n+2} (q, \theta_i(q), \mathbf{p}) \nonumber\\&
+\sum'_{\substack{g_1 + g_2 = g \\ J_1 \cup J_2 =  \mathbf{p}}} W^{g_1}_{|J_1|+1}(q, J_1) W^{g_2}_{|J_2|+1}(\theta_i(q), J_2) \Bigg) ,
\end{align}
which is equivalent to the original recursion of \cite{Eynard:2007,Eynard:2008}.

\subsection{Properties of the local recursion}

In this section we prove three lemmas about the $\mathcal{W}$ that were introduced in Definition \ref{d:curly}. These lemmas will be useful later on.

We start by proving an important property of the curly $\mathcal{W}$ defined in Definition \ref{d:curly}. We use the notation from subsection \ref{s:notation}.

\begin{lem}\label{l:curly1}
\begin{equation}
\mathcal{W}^g_{k,n}(\mathbf{q}; \mathbf{p}) = \mathcal{W}^{g-1}_{k-1,n+1} (\mathbf{q} \backslash \{q_k\}; q_k, \mathbf{p}) + \sum'_{\substack{g_1 + g_2 = g \\ J_1 \cup J_2 =  \mathbf{p}}} \mathcal{W}^{g_1}_{k-1, |J_1|}(\mathbf{q}  \backslash \{q_k\}; J_1) W^{g_2}_{|J_2|+1}(q_k, J_2).
\end{equation}
\end{lem}

\begin{proof}
By definition, the LHS is given by a sum over set partitions:
\begin{equation}
\mathcal{W}^g_{k,n}(\mathbf{q}; \mathbf{p}) = \sum_{\mu \in \text{Partitions}(\mathbf{q})} \sum'_{\substack{\sum_{i=1}^{\ell(\mu)} g_i = g + \ell(\mu) - k \\ \cup_{i=1}^{\ell(\mu)} J_i =  \mathbf{p}}} \left( \prod_{i=1}^{\ell(\mu)} W^{g_i}_{|\mu_i| + |J_i|} (\mu_i, J_i)  \right).
\end{equation}
The first term on the RHS, $\mathcal{W}^{g-1}_{k-1,n+1} (\mathbf{q} \backslash \{q_k\}; q_k, \mathbf{p})$, is also a sum over set partitions, but now of $\{q_1, \ldots, q_{k-1} \}$. $q_k$ is then distributed in all possible ways. This means that all terms in the summation over set partitions on the LHS that do not involve a subset of length one containing only $q_k$ are reproduced by the first term on the RHS.

As for partitions of $\mathbf{q}$ with a subset of length one given by $\{q_k\}$ itself, they are then reproduced by the explicit summation in the second term on the RHS:
\begin{equation}
\sum'_{\substack{g_1 + g_2 = g \\ J_1 \cup J_2 =  \mathbf{p}}} \mathcal{W}^{g_1}_{k-1, |J_1|}(\mathbf{q} \backslash \{q_k\}; J_1) W^{g_2}_{|J_2|+1}(q_k, J_2).
\end{equation}
\end{proof}

We can then use this property to relate $\mathcal{W}$ to contour integrals of higher order $\mathcal{W}$ with respect to the recursion kernel:

\begin{lem}\label{l:residue}
Let $C$ be a spectral curve. Construct the correlation functions $W^g_n$ for $C$ from the local recursion, Definition \ref{d:recursion}. Then, the $\mathcal{W}$ defined in Definition \ref{d:curly} satisfy:
\begin{equation}
\mathcal{W}^g_{2,n}(q_1,q_2; \mathbf{p}) = \frac{1}{2 \pi i} \sum_{i=1}^m \oint_{\Gamma_i(q)}  \left(\sum_{ \mathbf{d}_i'(q) }  K_{|\mathbf{d}_i'|+1}(q_1;q,\mathbf{d}_i'(q)) \mathcal{W}^g_{|\mathbf{d}_i'|+2, n}(q, \mathbf{d}_i'(q), q_2; \mathbf{p}) \right).\label{eq:residue}
\end{equation}
\end{lem}

\begin{proof}
We start by expanding the LHS:
\begin{equation}
\mathcal{W}^g_{2,n}(q_1,q_2; \mathbf{p}) = W^{g-1}_{n+2}(q_1, q_2, \mathbf{p}) + \sum'_{\substack{g_1 + g_2 = g \\ J_1 \cup J_2 =  \mathbf{p}}} W^{g_1}_{ |J_1|+1}(q_1,J_1) W^{g_2}_{|J_2|+1}(q_2, J_2).
\end{equation}
We assume that the correlation functions satisfy the local recursion. Therefore, we can substitute the recursion, Definition \ref{d:recursion},  for $W^{g-1}_{n+2}(q_1, q_2, \mathbf{p}) $ and $W^{g_1}_{ |J_1|+1}(q_1,J_1) $ in the RHS. We get:
\begin{align}
\mathcal{W}^g_{2,n}(q_1,q_2; \mathbf{p}) &=
 \frac{1}{2 \pi i} \sum_{i=1}^m \oint_{\Gamma_i(q)} \Bigg( \sum_{\mathbf{d}_i'(q) }  K_{|\mathbf{d}_i'|+1}(q_1;q,\mathbf{d}_i'(q)) \mathcal{W}^{g-1}_{|\mathbf{d}_i'|+1,n+1}(q, \mathbf{d}_i'(q); q_2,\mathbf{p} )\nonumber\\
&+  \sum'_{\substack{g_1 + g_2 = g \\ J_1 \cup J_2 =  \mathbf{p}}} \Bigg( \sum_{\mathbf{d}_i'(q) } K_{|\mathbf{d}_i'|+1}(q_1;q,\mathbf{d}_i'(q)) \mathcal{W}^{g_1}_{|\mathbf{d}_i'|+1,|J_1|}(q, \mathbf{d}_i'(q); J_1 )  \Bigg)W^{g_2}_{|J_2|+1}(q_2, J_2) \Bigg).
\end{align}
We factor out the summation over the subsets $\mathbf{d}_i'(q)$ and the recursion kernel. We get:
\begin{align}
\mathcal{W}^g_{2,n}(q_1,q_2; \mathbf{p}) =
 \frac{1}{2 \pi i} \sum_{i=1}^m \oint_{\Gamma_i(q)}& \Bigg(\sum_{ \mathbf{d}_i'(q) }  K_{|\mathbf{d}_i'|+1}(q_1;q,\mathbf{d}_i'(q)) \Bigg( \mathcal{W}^{g-1}_{|\mathbf{d}_i'|+1,n+1}(q, \mathbf{d}_i'(q); q_2,\mathbf{p} )\nonumber\\
& +  \sum'_{\substack{g_1 + g_2 = g \\ J_1 \cup J_2 =  \mathbf{p}}} \mathcal{W}^{g_1}_{|\mathbf{d}_i'|+1,|J_1|}(q, \mathbf{d}_i'(q); J_1 )W^{g_2}_{|J_2|+1}(q_2, J_2) \Bigg) \Bigg).
\end{align}
Using Lemma \ref{l:curly1}, it then follows that
\begin{equation}
\mathcal{W}^g_{2,n}(q_1,q_2; \mathbf{p}) =  \frac{1}{2 \pi i} \sum_{i=1}^m \oint_{\Gamma_i(q)} \Bigg( \sum_{\mathbf{d}_i'(q) }  K_{|\mathbf{d}_i'|+1}(q_1;q,\mathbf{d}_i'(q)) \mathcal{W}^g_{|\mathbf{d}_i'|+2, n}(q, \mathbf{d}_i'(q), q_2; \mathbf{p}) \Bigg),
\end{equation}
and the lemma is proved.
\end{proof}

\begin{rem}
Note that in the case of a curve where $\mathrm{d} x$ has only simple zeroes, at each $a_i \in R$ there is a unique non-trivial local deck transformation $\theta_i(q)$, hence the summation over subsets $\mathbf{d}_i'(q)$ collapses. In this particular case \eqref{eq:residue} becomes
\begin{equation}
\mathcal{W}^g_{2,n}(q_1,q_2; \mathbf{p}) = \frac{1}{2 \pi i} \sum_{i=1}^m \oint_{\Gamma_i(q)}  \left( K_{2}(q_1;q,\theta_i(q)) \mathcal{W}^g_{3, n}(q, \theta_i(q), q_2; \mathbf{p}) \right).
\end{equation}
\end{rem}

The previous lemma can be extended, by induction, to general curly $\mathcal{W}$:

\begin{lem}\label{l:residueAll}
Let $C$ be a spectral curve. Construct the correlation functions $W^g_n$ for $C$ from the local recursion, Definition \ref{d:recursion}. Then, the $\mathcal{W}$ defined in Definition \ref{d:curly} satisfy:
\begin{equation}
\mathcal{W}^g_{k,n}(\mathbf{q}; \mathbf{p})\nonumber\\ = \frac{1}{2 \pi i} \sum_{i=1}^m \oint_{\Gamma_i(q)} \left(\sum_{\mathbf{d}_i'(q)}  K_{|\mathbf{d}_i'|+1}(q_1;q,\mathbf{d}_i'(q)) \mathcal{W}^g_{|\mathbf{d}_i'|+k, n}(q, \mathbf{d}_i'(q),\mathbf{q} \backslash \{q_1\}; \mathbf{p}) \right).\label{eq:residueAll}
\end{equation}
\end{lem}

\begin{proof}
We prove this lemma by induction on $k$. The initial case $k=2$ is Lemma \ref{l:residue}. Now assume that \eqref{eq:residueAll} holds for all $k' < k$. We want to show that it then also holds for $k$. Using Lemma \ref{l:curly1}, we can expand the LHS as
\begin{equation}
\mathcal{W}^g_{k,n}(\mathbf{q}; \mathbf{p}) = \mathcal{W}^{g-1}_{k-1,n+1} (\mathbf{q} \backslash \{q_k\}; q_k, \mathbf{p})
+ \sum'_{\substack{g_1 + g_2 = g \\ J_1 \cup J_2 =  \mathbf{p}}} \mathcal{W}^{g_1}_{k-1, |J_1|}(\mathbf{q} \backslash \{q_k\}; J_1) W^{g_2}_{|J_2|+1}(q_k, J_2).
\end{equation}
Assuming that \eqref{eq:residueAll} holds for $k'=k-1$, we can use it to replace the curly $\mathcal{W}$'s:
\begin{gather}
\mathcal{W}^g_{k,n}(\mathbf{q}; \mathbf{p}) = \frac{1}{2 \pi i} \sum_{i=1}^m \oint_{\Gamma_i(q)} \Bigg( \sum_{ \mathbf{d}_i'(q)}  K_{|\mathbf{d}_i'|+1}(q_1;q,\mathbf{d}_i'(q)) \mathcal{W}^{g-1}_{|\mathbf{d}_i'|+k-1,n+1}(q, \mathbf{d}_i'(q), \mathbf{q} \backslash\{q_1,q_k\}; q_k,\mathbf{p} )\nonumber\\
+  \sum'_{\substack{g_1 + g_2 = g \\ J_1 \cup J_2 =  \mathbf{p}}} \Bigg( \sum_{\mathbf{d}_i'(q)} K_{|\mathbf{d}_i'|+1}(q_1;q,\mathbf{d}_i'(q)) \mathcal{W}^{g_1}_{|\mathbf{d}_i'|+k-1,|J_1|}(q, \mathbf{d}_i'(q), \mathbf{q} \backslash\{q_1,q_k\}; J_1 )  \Bigg)W^{g_2}_{|J_2|+1}(q_k, J_2) \Bigg).
\end{gather}
We factor out the summation over the subsets $\mathbf{d}_i'(q)$ and the recursion kernel. We get:
\begin{align}
\mathcal{W}^g_{k,n}(\mathbf{q}; \mathbf{p}) =\frac{1}{2 \pi i} \sum_{i=1}^m \oint_{\Gamma_i(q)} \sum_{ \mathbf{d}_i'(q) } & K_{|\mathbf{d}_i'|+1}(q_1;q,\mathbf{d}_i'(q)) \Bigg( \mathcal{W}^{g-1}_{|\mathbf{d}_i'|+k-1,n+1}(q, \mathbf{d}_i'(q),  \mathbf{q} \backslash\{q_1,q_k\}; q_k,\mathbf{p} )\nonumber\\
& +  \sum'_{\substack{g_1 + g_2 = g \\ J_1 \cup J_2 =  \mathbf{p}}} \mathcal{W}^{g_1}_{|\mathbf{d}'|+k-1,|J_1|}(q, \mathbf{d}_i', \mathbf{q} \backslash\{q_1,q_k\}; J_1 )W^{g_2}_{|J_2|+1}(q_k, J_2) \Bigg).
\end{align}
Then, by Lemma \ref{l:curly1}, we obtain
\begin{equation}
\mathcal{W}^g_{k,n}(\mathbf{q}; \mathbf{p}) =\frac{1}{2 \pi i} \sum_{i=1}^m \oint_{\Gamma_i(q)}\sum_{ \mathbf{d}_i'(q)}  K_{|\mathbf{d}_i'|+1}(q_1;q,\mathbf{d}_i'(q)) \Bigg(\mathcal{W}^{g}_{|\mathbf{d}_i'|+k, n}(q,\mathbf{d}_i'(q),\mathbf{q} \backslash\{q_1\}; \mathbf{p})\Bigg),
\end{equation}
and the lemma is proved.
\end{proof}

\begin{rem}
Consider again the case of a curve where $\mathrm{d} x$ has only simple zeroes, in which case at each $a_i \in R$ there is a unique non-trivial local deck transformation $\theta_i(q)$. Then \eqref{eq:residue} becomes
\begin{equation}
\mathcal{W}^g_{k,n}(\mathbf{q}; \mathbf{p}) =\frac{1}{2 \pi i} \sum_{i=1}^m \oint_{\Gamma_i(q)}  \Bigg( K_{2}(q_1;q,\theta_i(q))\mathcal{W}^{g}_{k+1, n}(q,\theta_i(q),\mathbf{q} \backslash\{q_1\}; \mathbf{p})\Bigg).
\end{equation}
\end{rem}

\section{A global topological recursion}

In the previous section we defined a recursion, Definition \ref{d:recursion}, based on \cite{BHLMR}, which is a generalization of the original recursion of \cite{Eynard:2007, Eynard:2008}. We named it ``local recursion'' for the following reason: in Definition \ref{d:recursion}, for each contour integral around $\Gamma_i(q)$, the integrand involves the local deck transformations generated by $\theta_i(q)$. Those are only defined locally on a disk $U_i$ around the ramification points $a_i \in S$. Hence the integrand is only defined locally on $U_i$. While this is perfectly fine in order to perform the integrals involved in the recursion, it may be desirable to rewrite the recursion in such a way that the integrand becomes a globally defined meromorphic differential form on $S$. 

One of the reason for defining a global version of the recursion is to study what happens to the correlation functions when ramification points collide. As an example, suppose that we consider the limit of a curve where two simple ramification points with a single common sheet collide to produce a single ramification of multiplicity $3$. The original correlation functions are produced by the local recursion for simple ramification points, that is, the original recursion of \cite{Eynard:2007, Eynard:2008}. After taking the limit however, the correlation functions are constructed from the local recursion in Definition \ref{d:recursion}, since there is now a double ramification point. The question is: are the correlation functions constructed from the curve after the limit equal to the limit of the correlation functions constructed on the original curve? In other words, does the local recursion Definition \ref{d:recursion} ``commute'' with taking limits of curves where ramification points collide?

This is particularly interesting because many properties of the correlation functions constructed from the original recursion for curves with simple ramification have been proved in \cite{Eynard:2007,Eynard:2008}. It would be desirable to study similar properties for the local recursion Definition \ref{d:recursion}, as studied in various examples in \cite{BHLMR}. However, the proofs become much more involved, because of the combinatorial nature of Definition \ref{d:recursion}. But if we know that the correlation functions constructed from Definition \ref{d:recursion} are in fact just appropriate limits of the correlation functions constructed from a curve with only simple ramification, then it is easy to see that most of the properties in \cite{Eynard:2007,Eynard:2008} carry through without modification to the recursion in Definition \ref{d:recursion}.

Unfortunately, it is difficult to study limits of correlation functions constructed from Definition \ref{d:recursion}, or the original recursion in \cite{Eynard:2007,Eynard:2008}. The problem is that the integrand in these recursion is only defined locally on the disks $U_i$ near the ramification points $a_i$; hence it makes it difficult to study what happens when two or more ramification points collide. To perform these limits we need to write the integrand as a meromorphic differential form at least over a neighborhood $U$ that includes both ramification points that collide in the particular limit considered. 

Instead of doing this case-by-case, in this section we propose a different approach. We will rewrite the recursion in a ``global way'', by which we mean that it will be written as a contour integral of a globally defined meromorphic differential form on $S$. Then we will show that locally on $U_i$ the integrand of our global recursion reduces to the local integrand of Definition \ref{d:recursion}. This then implies that the local recursion in Definition \ref{d:recursion} indeed ``commutes'' with taking limits of curves where ramification points collide, in the sense described above. Indeed, start with a curve, then rewrite the local recursion as a global recursion. Then you can take any particular limit you like, since the integrand is globally defined. Finally, after taking the limit you can rewrite the global recursion as a local recursion if you desire. It follows that the correlation functions then constructed are just limits of the original correlation functions constructed before taking the limit.

\subsection{Further assumptions}

To formulate a global recursion that reduces to Definition \ref{d:recursion} locally, we need to make two further assumptions on allowed spectral curves $C=(S,x,y)$.
\begin{defn}\label{d:acceptable}
Let $C=(S,x,y)$ be a spectral curve, $R = \{a_1, \ldots, a_n\}$ the zeroes of $\mathrm{d} x$, and $B = \{b_1,\ldots,b_m\} = \{x(a_1),\ldots,x(a_m)\}$ the corresponding branch points. We say that $C$ is \emph{acceptable} if it is such that:
\begin{itemize}
\item The branch points $b_i$, $i=1,\ldots,m$ are all distinct;
\item For any ramification point $a_j$, and any other point $p \in x^{-1}(b_j)$ in the preimage of the corresponding branch point $b_j = x(a_j)$, we must have that $y(a_j) \neq y(p)$.
\end{itemize}
\end{defn}

The first condition ensures that only one ramification point sits over a given branch point. We assume this condition for simplicity, but it seems rather harmless. We believe that our main Theorem \ref{t:main} still holds when more than one ramification point sit over a given branch point; the proof should go through with only minor modifications.

The second condition is however rather technical; what it means, in terms of the affine curve associated to $C$, is that for any ramification point $a_j \in S$, the point $(x(a_i),y(a_i))$ in the corresponding affine curve is not singular. For instance, Example \ref{ex:bad} is not an acceptable spectral cruve. This condition seems rather technical, but at the moment it is unclear to the authors how to get rid of it.

In the rest of the paper, unless specified, we will always assume that we are dealing only with acceptable spectral curves.

\subsection{Global meromorphic differentials}

The goal of this section is to rewrite the recursion as a contour integral over a globally defined meromorphic differential. For this purpose, we cannot use local deck transformations near the ramification points $a_i$. What we need to do instead is construct global meromorphic functions by summing over preimages of points under the branched covering $x: S \to \mathbb{C} \mathbb{P}^1$.

\subsubsection{Functions and differentials}

Let us start by considering meromorphic functions and differentials on $S$. We study the degree $d$ branched covering $x: S \to \mathbb{C} \mathbb{P}^1$. Given any meromorphic function $f(q)$ on $\mathbb{C} \mathbb{P}^1$, one can define the pullback $x^* f(p) = (f \circ x)(p)$, which is a meromorphic function on $S$. What we want to do now is a kind of reverse operation; given a meromorphic function on $S$, can we construct meromorphic functions on $\mathbb{C} \mathbb{P}^1$?

Let $f(p) \in K(S)$. A natural way of answering this question is by summing over preimages of a point in the base. More precisely, let $q \in \mathbb{C} \mathbb{P}^1$ be a generic point which is not a branch point of $x$. Then $x^{-1}(q)$ consists in precisely $d$ points $p_1, \ldots, p_n \in S$. Let $U_q$ be a small disk centered at $q$, and $U_{p_i}$ be small disks centered at the $p_i$. There is then a well defined biholomorphic map $\tau_i : U_d \to U_{p_i}$, for each preimage $p_i$ of $q$. 

Let us then define a function $\tilde{f}(t)$ on $U_q$ by summing over the pullback of $f$ under the maps $\tau_i$ (this is called the \emph{trace} of $f$):
\begin{equation}
\tilde{f}(t) = \sum_{i=1}^d \tau_i^* f(t) = \sum_{i=1}^d (f \circ \tau_i)(t).
\end{equation}
This is a well defined meromorphic function on $U_q$. In fact, it is meromorphic over the whole base $\mathbb{C} \mathbb{P}^1$ minus the branch points. Indeed, if we analytically continue $\tilde{f}(t)$ around paths in the base, all that can happen is that the $\tau_i^* f(t)$ get permuted due to monodromy, but the sum remains the same. Hence $\tilde{f}(t)$ is a well defined meromorphic function on the base minus the branch points. Finally, it is well known that it can be analytically continued to the whole base $\mathbb{C} \mathbb{P}^1$: this follows from Riemann's removable singularity theorem (see for instance \cite{Griffiths}). Thus $\tilde{f}(t)$ is a meromorphic function on $\mathbb{C} \mathbb{P}^1$.

Then we can pull it back to $S$ to obtain a new meromorphic function on $S$:
\begin{equation}
\hat{f}(p) = x^* \tilde{f}(p) = \tilde{f}(x(p)) = \sum_{i=1}^d f(\tau_i(x(p)).
\end{equation}
To stay sane, we will simplify the notation as $\tau_i(p) \coloneqq \tau_i(x(p))$. $\hat{f}(p)$ is meromorphic on $S$. But this is not quite the function we want yet. Now let $p \in S$, and choose the labeling of the sheets such that $\tau_{d}(p) = p$. Then
\begin{equation}
\hat{f}(p) =  \sum_{i=1}^{d-1} f(\tau_i(p)) + f(p).
\end{equation}
But $f(p)$ is of course meromorphic, hence we obtain that the function
\begin{equation}
F(p) := \hat{f}(p) - f(p)=  \sum_{i=1}^{d-1} f(\tau_i(p))
\end{equation}
is also a globally defined meromorphic function on $S$. Here we are summing over all preimages of $x(p)$ distinct from $p$.

What about differentials? Well the exact same construction can be applied to a meromorphic differential form $\omega$ on $S$. We can write $\omega(p) = f(p) \mathrm{d} g(p)$ for two meromorphic functions $f,g \in K(S)$. Given $\omega$, we construct a meromorphic differential $\tilde{\omega}(t)$ on $\mathbb{C} \mathbb{P}^1$ by summing over all preimages. Then we pull back to a meromorphic differential $\hat{\omega}(p)$ on $S$. By labeling the sheets appropriately, we single out the original meromorphic differential $\omega(p)$ from the summation, and we obtain a new meromorphic differential
\begin{equation}
\Omega(p) := \hat{\omega}(p) - \omega(p) = \sum_{i=1}^{d-1} \omega(\tau_i(p)) =  \sum_{i=1}^{d-1} f(\tau_i(p)) \mathrm{d} g(\tau_i(p)),
\end{equation}
where the sum is over all preimages of $x(p)$ distinct from $p$.

\subsubsection{Symmetric meromorphic differentials}

Now can we do a similar construction for symmetric meromorphic differentials on $S$ with poles along $R$? Let us start with a degree $2$ differential $\omega(p_1, p_2) \in \text{Sym}^2 H^0(S, K_S(* R) )$. Let $p \in S$, with $x(p) \in \mathbb{C} \mathbb{P}^1$. We construct the new object
\begin{equation}
\Omega(p) = \left( \sum_{i=1}^{d} \omega(p, \tau_i(p)) \right)- \omega(p,p),
\end{equation}
which is a well defined quadratic differential on $S$. Labeling the sheets such that $\tau_d(p) = p$, we get
\begin{equation}
\Omega(p) = \sum_{i=1}^{d-1} \omega(p, \tau_i(p)),
\end{equation}
where we are summing over all preimages of $x(p)$ distinct from $p$.

Note that we could have done the exact same construction for a degree $n$ differential, to obtain a new meromorphic differential:
\begin{equation}
\Omega(p,p_3,\ldots,p_n) = \sum_{i=1}^{d-1} \omega(p, \tau_i(p),p_3,\ldots,p_n).
\end{equation}

Now we can iterate the construction.  Consider the object
\begin{align}
\Omega(p_,p_4,\ldots,p_n) =& \left(\sum_{1 \leq \mu_1 < \mu_2 \leq d} \omega(p,\tau_{\mu_1}(p), \tau_{\mu_2}(p), p_4, \ldots, p_n)\right) \\
&- \left(\sum^{d-1}_{\mu_1=1} \omega(p,\tau_{\mu_1}(p),p, p_4, \ldots, p_n)\right).
\end{align}
The first term on the RHS is a well defined meromorphic differential by construction, because of symmetry of the differentials. The second term on the RHS is also a well defined meromorphic differential, as shown above. Hence, by labeling the sheets such that $\tau_d(p) = p$, we obtain that
\begin{equation}
\Omega(p_,p_4,\ldots,p_n) =\sum_{1 \leq \mu_1 < \mu_2 \leq d-1} \omega(p,\tau_{\mu_1}(p), \tau_{\mu_2}(p), p_4, \ldots, p_n)
\end{equation}
is a well defined meromorphic differential. Iterating, we finally obtain that for any $k$ and $n$, the object
\begin{equation}
 \Omega(p, p_{1}, \ldots, p_{k}) = \sum_{1 \leq \mu_1 < \mu_2 < \ldots < \mu_{n} \leq d-1} \omega(p, \tau_{\mu_1}(p), \ldots, \tau_{\mu_{n}}(p), p_{1}, \ldots, p_{k})
\end{equation}
is a well defined meromorphic differential.

\subsubsection{Differentials with poles on the diagonal}

In the previous subsection we considered symmetric meromorphic differentials that only have poles in each variable along $R \subset S$. What about differentials, such as the canonical bilinear differential, that have poles along the diagonals, \emph{i.e.} when two of the $p_i$'s coincide?

It is easy to see that the argument in the previous section goes through. One has to be careful when we specialize the $p_i$ to a given $p \in S$, to make sure that the result is a well defined meromorphic differential (and not identically infinity, for instance). But at every step that we specialized, we removed the poles along the diagonal; hence the result is a well defined differential. In other words,
\begin{equation}\label{eq:form}
 \Omega(p, p_{1}, \ldots, p_{k}) = \sum_{1 \leq \mu_1 < \mu_2 < \ldots < \mu_{n} \leq d-1} \omega(p, \tau_{\mu_1}(p), \ldots, \tau_{\mu_{n}}(p), p_{1}, \ldots, p_{k})
\end{equation}
is a well defined meromorphic differential, even if the original differential $\omega$ has poles along the diagonals.

\subsection{A global recursion}

We are now ready to define the global recursion. 

\subsubsection{Notation}

Let us first fix notation. Let $C=(S,x,y)$ be a spectral curve. We denote by $R = \{a_1, \ldots, a_m\}$ the set of zeroes of $\mathrm{d} x$, with corresponding multiplicities $r_i = \text{mult}_{a_i}(x)$. Let $d$ be the degree of the branched covering $x: S \to \mathbb{C} \mathbb{P}^1$. Let $p \in S$ be a point that is not in the preimage of a branch point of $x$; denote by $\tau_\mu(p):= \tau_\mu(x(p))$, $\mu=1,\ldots,d-1$ the preimages of $x(p)$ distinct from $p$. We introduce a vector notation $\bm{\tau}(p) = \{\tau_1(p), \ldots, \tau_{d-1}(p) \}$.

As before, to denote points on the Riemann surface $S$, we will use the letters $p$ and $q$. We will also use the set notation $\mathbf{p} = \{ p_1, \ldots, p_n \} \in S^n$ and $\mathbf{q}=\{ q_1,\ldots, q_k \} \in S^k$.

\subsubsection{The global recursion}

As usual, given a spectral curve $C$, we construct recursively an infinite tower of symmetric meromorphic differentials $\{ W_{n}^g \}$ with poles only along $R$. Let $W_{2}^0(p_1, p_2)$ be the canonical bilinear differential of the second kind on $S \times S$. 

We define the $\mathcal{W}$ as in Definition \ref{d:curly}, and the recursion kernel as in Definition \ref{d:kernel}.

We then define the integrand of the global recursion:

\begin{lem}\label{l:integrand}
Consider the expression
\begin{equation}\label{eq:exp}
\sum_{\bm{\tau}'(q)}  K_{|\bm{\tau}'|+1}(p_0;q,\bm{\tau}'(q)) \mathcal{W}^g_{|\bm{\tau}'|+1,n}(q, \bm{\tau}'(q); \mathbf{p} ),
\end{equation}
where the summation is over all non-empty subsets of $\bm{\tau}(p) = \{\tau_1(p), \ldots, \tau_{d-1}(p) \}$. Then it is a globally well defined meromorphic differential (in particular in $q$).
\end{lem}

\begin{proof}
By construction it is of course a well defined meromorphic differential in $p_0$ and $\mathbf{p}$. What we need to check is that it is also a well defined global meromorphic differential in $q$. But because of symmetry of the kernel, see Definition \ref{d:kernel}, and the $\mathcal{W}$, see Definition \ref{d:curly}, the summation over the subsets $\bm{\tau}'(q)$ produces an expression precisely of the form of \eqref{eq:form}. For instance, start with the differential
\begin{equation}
\omega(q_1,q_2, p_0, \mathbf{p}) = K_2(p_0; q_1, q_2) \mathcal{W}^g_{2,n}(q_1, q_2; \mathbf{p}),
\end{equation}
and symmetrize as in \eqref{eq:form}; the result is the sum over length one subsets $|\bm{\tau}'(q)|=1$ of $\bm{\tau}(q)$. Generally, let $\mathbf{q} =\{q_1, \ldots, q_k \}$ with $k \leq d$; start with
\begin{equation}
\omega(\mathbf{q}, p_0, \mathbf{p}) = K_k(p_0; \mathbf{q}) \mathcal{W}^g_{k,n}(\mathbf{q}; \mathbf{p}),
\end{equation}
and symmetrize as in \eqref{eq:form}; the result is the sum over subsets $|\bm{\tau}'(q)|=|\mathbf{q}|-1$ of $\bm{\tau}(q)$. Therefore, the expression \eqref{eq:exp} is a globally well defined meromorphic differential in $q$.
\end{proof}

We then define the global recursion:
\begin{defn}\label{d:recursionglobal}
Let $C$ be a spectral curve. We uniquely construct degree $n$ symmetric meromorphic differentials $W_n^g$ on $S$ with poles along $R$ through the \emph{global recursion}:
\begin{equation}\label{e:recursionglobal}
W^g_{n+1}(p_0, \mathbf{p}) = \frac{1}{2 \pi i} \oint_{\Gamma(q)} \left(\sum_{\bm{\tau}'(q)}  K_{|\bm{\tau}'|+1}(p_0;q,\bm{\tau}'(q)) \mathcal{W}^g_{|\bm{\tau}'|+1,n}(q, \bm{\tau}'(q); \mathbf{p} ) \right).
\end{equation}
The contour integral here is in $q$, over the union of small circles $\Gamma_i(q)$ around each ramification point.
\end{defn}

\begin{rem}
Note that Definition \ref{d:recursionglobal} looks very similar to Definition \ref{d:recursion}, but it is in fact quite different. First, the integrand in Definition \ref{d:recursionglobal} is now a globally defined meromorphic differential, which is why we can write the contour integral as a single contour integral over the union of small circles; in Definition \ref{d:recursion} the integrand was only defined locally on a disk around the ramification points. 

Also, the summation in \ref{d:recursionglobal} is over the non-empty subsets $\bm{\tau}'(q)$ of the $d-1$ preimages of $x(q)$ distinct from $q$, instead of being a summation over non-empty subsets of the local deck transformations at a ramification point. The two are very different. For instance, given a spectral curve with only simple ramification points, the summation in Definition \ref{d:recursion} collapses, and only terms of the form $K_2 \mathcal{W}^g_{2,n}$ enter into the recursion. However, in Definition \ref{d:recursionglobal} the integrand depends on the degree $d$ of the branched covering $x: S \to \mathbb{C} \mathbb{P}^1$, not on the multiplicity of the ramification points. Thus, if $x$ is a degree $d$ branched covering with only simple ramification points, in Definition \ref{d:recursionglobal} we still have all terms from $K_2 \mathcal{W}^g_{2,n}$ to $K_d \mathcal{W}^g_{d,n}$. This is the price to pay for having a globally defined integrand. 

To summarize, in Definition \ref{d:recursion} the integrand is only defined locally near the $a_i$, and depends on the multiplicity of $a_i$; in \ref{d:recursionglobal} the integrand is globally defined, and depends only on the degree $d$ of the branched covering.
\end{rem}

\subsection{Localizing the global recursion}

In the previous subsection we proposed a ``global'' recursion, where the integrand is globally defined and depends only on the degree of the branched covering. What we now need to show is that if we analyze the integrand locally in a disk around each $a_i$, then Definition \ref{d:recursionglobal} reduces to Definition \ref{d:recursion} for the local recursion. In particular, for spectral curves with only simple ramification, it reduces to the original recursion of \cite{Eynard:2007,Eynard:2008}.

\begin{thm}\label{t:main}
Let $C$ be a spectral curve. Construct the correlation functions $W^g_n$ on $S$ through the global recursion Definition \ref{d:recursionglobal}. Then the $W^g_n$ also satisfy the local recursion Definition \ref{d:recursion}.
\end{thm}

We will prove this theorem by induction. Let us start by proving the theorem for $W^0_3$ and $W^1_1$, the initial cases of the induction.

\begin{lem}\label{l:03}
Let $C$ be a spectral curve. Construct the correlation function $W^0_3$ on $S$ through the global recursion Definition \ref{d:recursionglobal}. Then $W^0_3$ also satisfies the local recursion Definition \ref{d:recursion}.
\end{lem}

\begin{proof}
First, note that for $(g,n) = (0,3)$, all $\mathcal{W}^0_{k,2}=0$ for $k \geq 3$. Thus, Definition \ref{d:recursionglobal} reduces to
\begin{equation}
W^0_{3}(p_0, p_1,p_2) = \frac{1}{2 \pi i} \oint_{\Gamma(q)} \left(\sum_{\mu=1}^{d-1}  K_{2}(p_0;q,\tau_\mu(q)) \mathcal{W}^0_{2,2}(q,\tau_\mu(q); p_1,p_2 ) \right).
\end{equation}
Now let us evaluate the contour integral around the circle $\Gamma_1(q)$ near $a_1$. Assume that it has multiplicity $r_1$. Then there are $r_1$ sheets that collide at $a_1$. Let $U_1$ be a small disk around $a_1$, and $q \in U_1$. Without loss of generality we suppose that the labeling is such that the $r_1-1$ other sheets that collide at $a_1$ are $\tau_1(q), \ldots, \tau_{r_1-1}(q)$. Thus for $q \in U_1$, we have $\tau_1(q), \ldots, \tau_{r_1-1}(q) \in U_1$, while $\tau_{r_1}(q), \ldots, \tau_{d-1}(q)$ correspond to the other preimages of $x(q)$.

Now if we restrict ourselves to $q \in U_1$, then we can split the integrand into a number of well defined meromorphic functions on $U_1$. Namely, the sum
\begin{equation}\label{eq:firstterm}
\sum_{\mu=1}^{r_1-1}  K_{2}(p_0;q,\tau_\mu(q)) \mathcal{W}^0_{2,2}(q,\tau_\mu(q); p_1,p_2 )
\end{equation}
is a well defined meromorphic funtion (since monodromy can only permute these sheets), while the terms
\begin{equation}
K_{2}(p_0;q,\tau_\nu(q)) \mathcal{W}^0_{2,2}(q,\tau_\nu(q); p_1,p_2 )
\end{equation}
for $\nu =r_1, \ldots, d-1$ are individually well defined on $U_1$. In fact, the integral of \eqref{eq:firstterm} around $\Gamma_1(q)$
\begin{equation}
\frac{1}{2 \pi i} \oint_{\Gamma_1(q)} \sum_{\mu=1}^{r_1-1}  K_{2}(p_0;q,\tau_\mu(q)) \mathcal{W}^0_{2,2}(q,\tau_\mu(q); p_1,p_2 )
\end{equation}
is precisely the corresponding term in the local recursion Definition \ref{d:recursion}, since the $\tau_\mu(q)$ restricted to $U_1$ are precisely equal to the local deck transformations $\theta_1^\mu (q)$. Thus what we need to show is that
\begin{equation}\label{eq:contour}
\frac{1}{2 \pi i} \oint_{\Gamma_1(q)} K_{2}(p_0;q,\tau_\nu(q)) \mathcal{W}^0_{2,2}(q,\tau_\nu(q); p_1,p_2 ) = 0,
\end{equation}
for $\nu = r_1, \ldots, d-1$.

First, note that $K_{2}(p_0;q,\tau_\nu(q))$ has a pole of order $r_1-1$ at $a_1$. Indeed, recall that
\begin{equation}
K_{2}(p_0;q,\tau_\nu(q)) = - \frac{dS_{q,0}(p_0)}{ (y(q) - y(\tau_\nu(q)) \mathrm{d} x(q)}.
\end{equation}
Since we assume that $y(a_1) \neq y(\tau_\nu(a_1))$ for any $\nu=r_1,\ldots,d-1$ (see Definition \ref{d:acceptable}), the only pole for $q$ on a small disk $U_1$ around $a_1$ is the pole of order $r_1-1$ in $\mathrm{d} x(q)$.

Then, note that $\mathcal{W}^0_{2,2}(q,\tau_\nu(q); p_1,p_2 ) $ has a zero of order $r_1-1$ at $a_1$. Recall that
\begin{equation}
\mathcal{W}^0_{2,2}(q,\tau_\nu(q); p_1,p_2 )  = W^0_2(q, p_1) W^0_2(\tau_\nu(q),p_2) + W^0_2(q, p_2) W^0_2(\tau_\nu(q),p_1),
\end{equation}
which clearly has no pole at $q=a_1$. Moreover, the map $\tau_\nu: V_1 \to U_{1,\nu}$ is one-to-one, where $V_1$ is a disk around $x(a_1)$ and $U_{1,\nu}$ is a disk around the preimage $\tau_\nu(x(a_1))$, since we assume that no other ramification point sits over $x(a_1)$. Thus the differential $\mathrm{d} \tau_\nu(p) = \mathrm{d} \tau_\nu(x(p))$ behaves just like $\mathrm{d} x(p)$ on $U_1$, namely, it has a zero of order $r_1-1$ at $a_1$.

Putting this together, the integrand, for $\mu=r_1, \ldots, d-1$, is regular at $q=a_1$, and the contour integral \eqref{eq:contour} is zero.

The same analysis of course holds for other ramification points $a_i$ (with appropriate identification of the sheets meetings at $a_i$). Therefore, the global recursion for $W^0_3$ is precisely equal to the local recursion.
\end{proof}

Consider now the second initial case:

\begin{lem}\label{l:11}
Let $C$ be a spectral curve. Construct the correlation function $W^1_1$ on $S$ through the global recursion Definition \ref{d:recursionglobal}. Then $W^1_1$ also satisfies the local recursion Definition \ref{d:recursion}.
\end{lem}

\begin{proof}
The proof is very similar. Again, for $(g,n) = (1,1)$, only $\mathcal{W}^1_{2,0}$ is non-zero. The global recursion becomes
\begin{equation}
W^1_{1}(p_0) = \frac{1}{2 \pi i} \oint_{\Gamma(q)} \left(\sum_{\mu=1}^{d-1}  K_{2}(p_0;q,\tau_\mu(q)) \mathcal{W}^1_{2,0}(q,\tau_\mu(q) ) \right).
\end{equation}
Consider the contour $\Gamma_1(q)$. Restrict $q$ to a disk $U_1$ around $a_1$. Label the sheets such that $q$ and the $\tau_\mu(q)$ for $\mu=1,\ldots,r_1-1$ correspond to the sheets meetings at $a_1$. Then the sum over these terms reproduces the term from the local recursion.

We thus want to evaluate the contour integral
\begin{equation}\label{eq:cc}
\oint_{\Gamma_1(q)}  K_{2}(p_0;q,\tau_\nu(q)) \mathcal{W}^1_{2,0}(q,\tau_\nu(q) ) ,
\end{equation}
for $\nu=r_1,\ldots,d-1$. Again, $K_{2}(p_0;q,\tau_\nu(q))$ has a pole of order $r_1-1$ at $a_1$, since for acceptable spectral curves $y(a_1) \neq y(\tau_\nu(a_1))$. Recall that
\begin{equation}
 \mathcal{W}^1_{2,0}(q,\tau_\nu(q) ) = W^0_2(q,\tau_\nu(q)),
\end{equation}
which is regular at $q=a_1$, since $\tau_\nu(q)$ corresponds to a preimage of $x(a_1)$ distinct from $a_1$. In fact, it has a zero of order $r_1-1$, due to the differential $\mathrm{d} \tau_\nu(q)$. As a result, the contour integral \eqref{eq:cc} vanishes for $\nu=r_1, \ldots, d-1$. 

We redo the analysis for other ramification points, and we obtain that the global recursion reduces to the local recursion for $W^1_1$.
\end{proof}

We are now ready to prove Theorem \ref{t:main} by induction.

\begin{proof}[Proof of Theorem \ref{t:main}]

We assume that the theorem is valid for all correlation functions $W^{g'}_{n'+1}$ with $2g'-2+n' < 2g-2+n$. The initial cases are $(g',n'+1)=(0,3)$ and $(g',n'+1) = (1,1)$, which were proved in the two previous lemmas.

Consider the global recursion for $W^g_{n+1}$:
\begin{equation}\label{eq:gl}
W^g_{n+1}(p_0, \mathbf{p}) = \frac{1}{2 \pi i} \oint_{\Gamma(q)} \left(\sum_{\bm{\tau}'(q)}  K_{|\bm{\tau}'|+1}(p_0;q,\bm{\tau}'(q)) \mathcal{W}^g_{|\bm{\tau}'|+1,n}(q, \bm{\tau}'(q); \mathbf{p} ) \right).
\end{equation}
Let us focus on the contour integral around $\Gamma_1(q)$. Let $q \in U_1$ near $a_1$, which has multiplicity $r_1$. Label the sheets such that $q$ and the $\tau_\mu(q)$ for $\mu=1,\ldots,r_1-1$ correspond to the sheets meetings at $a_1$.

We split the subsets $\bm{\tau}'(q)$ of $\{\tau_1(q), \ldots, \tau_{d-1}(q) \}$ in three different groups. First, write $\bm{\alpha}(q)$ for subsets of $\{\tau_1(q), \ldots, \tau_{r_1-1}(q)\}$, and $\bm{\beta}(q)$ for subsets of $\{ \tau_{r_1}(q), \ldots, \tau_{d-1}(q)\}$. Finally, write $\bm{\gamma}(q)$ for subsets that involve at least one sheet in both $\{\tau_1(q), \ldots, \tau_{r_1-1}(q)\}$ and $\{ \tau_{r_1}(q), \ldots, \tau_{d-1}(q)\}$. The sum over all $\bm{\alpha}(q)$, $\bm{\beta}(q)$ and $\bm{\gamma}(q)$ is clearly equivalent to the sum over all $\bm{\tau}'(q)$.

As in the proof of the previous lemmas, we note that if we restrict $q$ to $U_1$, then the sum over the subsets $\bm{\alpha}(q)$ provides a well defined meromorphic function on $U_1$ (since monodromy can only permute sheets in there). The sum over all ``mixed'' subsets $\bm{\gamma}(q)$ also provides a well defined meromorphic function, and so does the sum over subsets $\bm{\beta}(q)$.

Consider first the sum over the subsets $\bm{\alpha}(q)$:
\begin{equation}
W^g_{n+1}(p_0, \mathbf{p}) = \frac{1}{2 \pi i} \oint_{\Gamma_1(q)} \left(\sum_{\bm{\alpha}(q)}  K_{|\bm{\alpha}|+1}(p_0;q,\bm{\alpha}(q)) \mathcal{W}^g_{|\bm{\alpha}|+1,n}(q,\bm{\alpha}(q); \mathbf{p} ) \right).
\end{equation}
Restricted locally on $U_1$, the $\tau_1(q), \ldots, \tau_{r_1-1}(q)$ become the local deck transformations, and the expression above is precisely the expression obtained from the local recursion. Thus what we need to show is that the sum over the subsets $\bm{\beta}(q)$ and $\bm{\gamma}(q)$ cancel.

Consider now the sum over subsets $\bm{\beta}(q)$ of $\{ \tau_{r_1}(q), \ldots, \tau_{d-1}(q)\}$:
\begin{equation}\label{eq:toev}
 \oint_{\Gamma_1(q)} \left(\sum_{\bm{\beta}(q)}  K_{|\bm{\beta}|+1}(p_0;q,\bm{\beta}(q)) \mathcal{W}^g_{|\bm{\beta}|+1,n}(q,\bm{\beta}(q); \mathbf{p} ) \right).
\end{equation}
To evaluate this contour integral, we use the induction hypothesis. Since we assume that the theorem is valid for all correlation functions $W^{g'}_{n'+1}$ with $2g'-2+n' < 2g-2+n$, we can assume that the correlation functions in the $\mathcal{W}^g_{|\bm{\beta}|+1,n}(q,\bm{\beta}(q); \mathbf{p} )$ have been obtained through the local recursion Definition \ref{d:recursion}. In particular, the $\mathcal{W}^g_{|\bm{\beta}|+1,n}(q,\bm{\beta}(q); \mathbf{p} ) $ satisfy Lemma \ref{l:residueAll}. That is,
\begin{equation}
\mathcal{W}^g_{|\bm{\beta}|+1,n}(q,\bm{\beta}(q); \mathbf{p})\nonumber\\ = \frac{1}{2 \pi i} \sum_{i=1}^m \oint_{\Gamma_i(s)} \left(\sum_{\mathbf{d}_i'(s)}  K_{|\mathbf{d}_i'|+1}(q;s,\mathbf{d}_i'(s)) \mathcal{W}^g_{|\mathbf{d}_i'|+k, n}(s, \mathbf{d}_i'(s),\bm{\beta}(q); \mathbf{p}) \right),
\end{equation}
where we are using the local notation of Definition \ref{d:recursion}. Thus \eqref{eq:toev} becomes
\begin{gather}
 \frac{1}{2 \pi i}\sum_{i=1}^m  \oint_{\Gamma_1(q)}\oint_{\Gamma_i(s)}\Bigg(\sum_{\bm{\beta}(q)}  \sum_{\mathbf{d}_i'(s)}   K_{|\bm{\beta}|+1}(p_0;q,\bm{\beta}(q))K_{|\mathbf{d}_i'|+1}(q;s,\mathbf{d}_i'(s)) \nonumber\\ \mathcal{W}^g_{|\mathbf{d}_i'|+k, n}(s, \mathbf{d}_i'(s),\bm{\beta}(q); \mathbf{p})
 \Bigg).\label{eq:tocomp}
\end{gather}
This iterated contour integral is defined as follows. The contour $\Gamma_1(q)$ is a circle centered around $a_1$ of radius $\epsilon_1$. As for $\Gamma_i(s)$, they are also circles centered around $a_i$ of radius $\epsilon_i'$. For the case $i=1$, we must require that $\epsilon_1' < \epsilon_1$ to make sense of the iterated contour integral.

To evaluate this expression, we want to perform the contour integral in $q$ first. For all the contours with $i \neq 1$, we can do that without problem, since the contours are around distinct ramification points. However, for $i=1$ we need to be careful. If we perform the integration over $q$ first, we are integrating over a circle of radius $\epsilon_1$ around $a_1$, with $\epsilon_1 > \epsilon_1'$. Looking at the integrand, this means that we will also pick up a residue at the pole at $q=s$. 

Now we claim that the integrand has no pole at $q=a_1$. First, for acceptable spectral curves, the term 
\begin{equation}
K_{|\bm{\beta}|+1}(p_0;q,\bm{\beta}(q))
\end{equation}
has a pole of order $(r_1-1) |\bm{\beta}(q)|$ coming from the factor of $\mathrm{d} x^{|\bm{\beta}(q)|}$ in the denominator. Second, 
\begin{equation}
K_{|\mathbf{d}_i'|+1}(q;s,\mathbf{d}_i'(s))
\end{equation}
is regular at $q=a_1$, since its only poles in $q$ are at $q=s$ and $q=o$ where $o$ is an arbitrary base point (which we assume to be not in $U_1$). Finally, we claim that
\begin{equation}
\mathcal{W}^g_{|\mathbf{d}_i'|+k, n}(s, \mathbf{d}_i'(s),\bm{\beta}(q); \mathbf{p})
\end{equation}
has a zero of order $(r_1-1) |\bm{\beta}(q)|$ at $q=a_1$. First, it is clear that it is regular; the only dependence in $q$ comes from the sheets in $\bm{\beta}(q)$, which do not meet at $a_1$. By assumption, there is no other ramification point over $x(a_1)$, and the only poles of the $W^g_n$ are at the ramification points, hence the term must be regular at $q=a_1$. Moreover, each $\tau_\nu(q)$ in $\bm{\beta}(q)$ comes with its differential $\mathrm{d} \tau_\nu(q)$, which has a zero of order $(r_1-1)$ at $q=a_1$. Thus we obtain a zero of order $(r_1-1) |\bm{\beta}(q)|$.

As a result, the integrand is regular at $q=a_1$. Therefore the contour integral over $\Gamma_1(q)$ in \eqref{eq:tocomp} will pick up the residue at $q=s$ for the term $i=1$ in the summation over $i$. That is, \eqref{eq:tocomp} becomes
\begin{gather}
\oint_{\Gamma_1(s)} \underset{q=s}{\text{Res}}\Bigg(\sum_{\bm{\beta}(q)}  \sum_{\bm{\alpha}(s)}   K_{|\bm{\beta}|+1}(p_0;q,\bm{\beta}(q))K_{|\bm{\alpha}|+1}(q;s,\bm{\alpha}(s))  \mathcal{W}^g_{|\bm{\alpha}|+k, n}(s, \bm{\alpha}(s),\bm{\beta}(q); \mathbf{p})
 \Bigg),
\end{gather}
where we used the fact that on $U_1$ the local deck transformations $\mathbf{d}_1'(s)$ are just the sheets $\bm{\alpha}(s)$. Now the only pole at $q=s$ is in
\begin{equation}
K_{|\bm{\alpha}|+1}(q;s,\bm{\alpha}(s)),
\end{equation}
where it enters in $d S_{s,o}(q)$; it has a simple pole at $q=s$ with residue $+1$. Therefore, taking the residue at $q=s$ amounts to replacing
\begin{equation}
K_{|\bm{\beta}|+1}(p_0;q,\bm{\beta}(q))K_{|\bm{\alpha}|+1}(q;s,\bm{\alpha}(s)) 
\end{equation}
by
\begin{equation}
- K_{|\bm{\alpha}| + |\bm{\beta}|+1}(p_0;s,\bm{\alpha}(s),\bm{\beta}(s)), 
\end{equation}
and evaluating the rest of the expression at $q=s$. Thus \eqref{eq:tocomp} finally becomes
\begin{gather}
- \oint_{\Gamma_1(s)} \Bigg(\sum_{\bm{\beta}(s)}  \sum_{\bm{\alpha}(s)} K_{|\bm{\alpha}| + |\bm{\beta}|+1}(p_0;s,\bm{\alpha}(s),\bm{\beta}(s))\mathcal{W}^g_{|\bm{\alpha}|+k, n}(s, \bm{\alpha}(s),\bm{\beta}(s); \mathbf{p})
 \Bigg).
\end{gather}
But the sum here is over non-empty subsets $\bm{\alpha}(s)$ and $\bm{\beta}(s)$; in other words, it is precisely equal to the sum over the ``mixed'' subsets $\bm{\gamma}(s)$ that we defined earlier. Thus it can be rewritten as a single sum:
\begin{equation}
- \oint_{\Gamma_1(s)} \Bigg(\sum_{\bm{\gamma}(s)}  K_{|\bm{\gamma}| +1}(p_0;s,\bm{\gamma}(s))\mathcal{W}^g_{|\bm{\gamma}|+k, n}(s, \bm{\gamma}(s); \mathbf{p})
 \Bigg).
\end{equation}
Because of the minus sign, it precisely cancels with the corresponding sum coming from the global recursion \eqref{eq:gl}! As a result, the only terms that remain for the contour integral over $\Gamma_1(q)$ in \eqref{eq:gl} are the terms coming from the summation over the subset $\bm{\alpha}(q)$, which reproduce precisely the terms in the local recursion.

We can of course do an identical analysis for the other ramification points $a_i$; the final result is that for each ramification point $a_i$, the contour integral over $\Gamma_i(q)$ in \eqref{eq:gl} becomes precisely the corresponding contour integral in the local recursion. Therefore, the correlation functions $W^g_n$ constructed from the global recursion are precisely the same as those constructed from the local recursion, and the theorem is proved.
\end{proof}

\subsection{Limits of spectral curves}

This is all nice, but what is the use of the global recursion? After all, concretely to perform the contour integral we may as well work locally; so why bother writing a globally defined integrand?

The original goal was to understand the relation between the local recursion Definition \ref{d:recursion} and the original recursion of \cite{Eynard:2007,Eynard:2008}. In particular, given a spectral curve with only simple ramification points, one can consider a particular limit of the curve where two of the ramification points with a common sheet collide, and one would end up with a curve with simple ramification points and one double ramification points. The question was: would the correlation function constructed from Definition \ref{d:recursion} for the limiting curve be equal to the limit of the correlation functions constructed from the curve with only simple ramification?

We now have a direct answer; just go through the global recursion! Start with the recursion of \cite{Eynard:2007,Eynard:2008} for the curve with only simple ramification. We know from Theorem \ref{t:main} that it is equivalent to the global recursion, Definition \ref{d:recursionglobal}, for the same curve. But then, the integrand of the global recursion is globally defined, and only depends on the degree of the covering; it does not depend on the multiplicity of the ramification points. Thus we can very well take the limit explicitly in the integrand. Finally, after taking the limit, Theorem \ref{t:main} tells us that the recursion is equivalent to the local recursion, Definition \ref{d:recursion}, for the limiting curve with a double ramification point. Hence the answer to the question is affirmative; the limit of the original correlation functions will be equal to the correlation functions constructed from the limiting curve.

In fact, through the global recursion we can analyze any type of collision, not only two simple ramification points producing a double ramification point. It follows that the limit of the correlation functions under any kind of collision process will always be equal to the correlation functions obtained from the limiting curve.

We can use this knowledge to prove that the correlation functions constructed from the local recursion, Definition \ref{d:recursion}, or from the global recursion, Definition \ref{d:recursionglobal}, satisfy most of the same properties as the original correlation functions of \cite{Eynard:2007,Eynard:2008}: this is what we do in the next section.

\subsubsection{Extra assumptions}

Before we do that, let us mention a few more things about the extra assumptions on the spectral curve that were made in Definition \ref{d:acceptable}. The local recursion in Definition \ref{d:recursion} does not require these assumptions on the spectral curve; hence one may ask whether it is possible to consider limits of colliding ramification points, without requiring these extra assumptions.

The reasons why we needed to make these assumptions was to study the local limit of the global recursion defined in Definition \ref{d:recursionglobal}. But in fact, to study the limit of two colliding ramification points, we do not necessarily need to go through the global recursion. All we need to do is rewrite the local recursion in a way that is ``global'' only on an open set $U \subset S$ that contains both ramification points $a_1$ and $a_2$ that collide in the limit.

So we can formulate a ``mixed'' recursion, where we start with the local recursion, and only ``globalize'' part of it, by replacing the sums over deck transformations near $a_1$ and $a_2$ by sums over the sheets that meet at $a_1$ and $a_2$, along the lines of the previous subsections. The result will be an integrand that is meromorphic on $U$. We can always do that, as long as the assumptions in Definition \ref{d:acceptable} are \emph{satisfied locally on $U$} (which is of course weaker than assuming that they are satisfied globally on $S$). That is, we require that the two branch points be distinct, $b_1 = x(a_1) \neq x(a_2) = b_2$, and that for $a_1 \in U$, any other point $p \in x^{-1}(b_1)$ that is also in $U$ must satisfy $y(p) \neq y(a_1)$, and similarly for $a_2 \in U$. 

Under these local assumptions, by the steps of the previous subsections,  the mixed recursion reduces to the local version before and after the collision. This proves that the correlation functions constructed from the local recursion after the collision are equal to the limits of the correlation functions constructed from the local recursion before the collision, without having to assume that the requirements of Definition \ref{d:acceptable} are satisfied globally on $S$.

\subsubsection{Spectral curves in $\mathbb{C}^*$}

For many applications of the topological recursion in enumerative geometry, such as in mirror symmetry, topological string theory and knot theory, spectral curves are given in a form slightly different than what we have studied so far. They are usually given as an algebraic curve
\begin{equation}
\{ P(x,y) = 0 \} \subset (\mathbb{C}^*)^2,
\end{equation}
where $P(x,y)$ is a polynomial. When we defined a spectral curve in Definition \ref{d:spectral} as a triple $(S,x,y)$, we mentioned that the meromorphic functions $x$ and $y$ satisfy an equation $P(x,y) = 0$, which can be considered as an affine curve whose normalization is $S$. The difference here is that we would like to consider an affine curve with punctures; that is, we require that $x$ and $y$ live in $\mathbb{C}^*$.

How can we implement that in the recursion? Perhaps the easiest way is to keep the definition of a spectral curve, Definition \ref{d:spectral}, as it is; as a triple $(S,x,y)$, with $S$ a Torelli marked compact Riemann surface, and $x$ and $y$ meromorphic functions on $S$. But then we need to change two things in the formulation of the recursion:
\begin{itemize}
\item We replace $R$ by the zeroes of $\mathrm{d} x/x$ instead of the zeroes of $\mathrm{d} x$. The zeroes of $\mathrm{d} x/x$ are a subset of the zeroes of $\mathrm{d} x$; the missing zeroes are precisely those that are also zeroes of $x$, \emph{i.e.} punctures of the curve.
\item We replace the recursion kernel, Definition \ref{d:kernel}, by:
\begin{equation}
K_k(p_0;\mathbf{q}) = - \frac{dS_{q_1,o}(p_0)}{\prod_{i=2}^{k} \omega(q_1, q_i)},
\end{equation}
with
\begin{equation}
\omega (q_1,q_2) = \left( \log y(q_1) - \log y(q_2) \right) \frac{\mathrm{d} x(q_1)}{x(q_1)}.
\end{equation}
\end{itemize}

Can we then study colliding ramification points in this setup? $x$ has a finite number of sheets, just as before, hence we can still sum over sheets to define the integrand in Lemma \ref{l:integrand}. However, because of the $\log$'s in the recursion kernel, the expression will \emph{not} be globally meromorphic on $S$.

However, the only points $q\in S$ where the integrand in Lemma \ref{l:integrand} has a logarithmic singularity are the points where either $y(q) = 0,\infty$, or $y(\tau_i(q)) = 0,\infty$ for any other point $\tau_i(q) \in x^{-1} (x(q))$. Let $Q \subset S$ denote the set of points where the integrand has logarithmic singularities. We can then introduce cuts $C_k$ joining these points pairwise; the integrand is then meromorphic on $S \setminus \cup C_k$, \emph{i.e.} on the compact Riemann surface $S$ minus the cuts.

Is this sufficient to carry the steps in the previous subsection to show that the ``global'' recursion reduces to the local recursion? (Note that ``global'' here only means meromorphic on $S$ minus the cuts.) It is easy to see that the argument will go through if the integrand in Lemma \ref{l:integrand} is meromorphic on an open set $U$ that contains all the ramification points ($U$ does not have to be the whole compact $S$). Therefore, what we must show is that there exists an open set $U \subset S$ that contains all ramification points, but no cuts $C_k$.

Such an open set $U$ will exist if we require that for each zero $a_i \in S$ of $\mathrm{d} x/x$, all preimages $\tau_k(a_i) \in x^{-1}(x(a_i))$ are such that $y(\tau_k(a_i)) \neq 0,\infty$. Under this assumption,  we can define an open set $U$ that contains all ramification points, but no cuts $C_k$. Then it follows that the ``global'' recursion reduces to the local recursion, and all the conclusions of the previous subsections apply to this case as well.

\begin{rem}
Note that the extra requirement above is necessarily satisfied if we require that all punctures of the curve correspond to zeros and poles of $x$. In this case, if it happens that $y(\tau_k(a_i))=0$ or $\infty$ for a given sheet, then it is a puncture of the curve, hence $x(\tau_k(a_i)) = x(a_i) = 0$ or $\infty$, which is a contradiction with the definition of $a_i$ as a zero of $\mathrm{d} x/x$. Therefore it follows that $y(\tau_k(a_i)) \neq 0,\infty$ for all sheets, and the open set $U$ exists. However, note that requiring that all punctures of the curve are at zeros and poles of $x$ is a stronger condition than the requirement above.
\end{rem}

\begin{rem}
Finally, note that just as in the previous subsubsection, we could formulate this requirement only locally. If we consider a collision between two ramification points $a_1$ and $a_2$, then we could replace the intermediate step of constructing the global recursion by a ``mixed'' recursion, which would be meromorphic only on an open set $U'$ containing $a_1$ and $a_2$ but no cuts. Then, we would only need to require that $y(\tau_k(a_1)) \neq 0,\infty$ and $y(\tau_k(a_2)) \neq 0,\infty$, and only for the sheets that meet at $a_1$ and $a_2$.
\end{rem}

\section{Properties of the correlation functions}

It was shown that the correlation functions constructed from the original recursion of \cite{Eynard:2007,Eynard:2008} satisfy many interesting properties. An obvious question is whether the correlation functions constructed from the local recursion proposed in \cite{BHLMR}, and the global recursion studied in this paper, satisfy the same properties. It was checked in \cite{BHLMR} that those properties hold in various examples, but a proof was not given. This is what we study in this section.

\subsection{Symmetry}

Perhaps the most fundamental property of the original correlation functions $W^g_n(\mathbf{p})$ is that they are fully symmetric under permutations of the arguments. Symmetry under permutations of the arguments $\{p_2, \ldots, p_n\}$ is built in the recursion, but symmetry under permutations involving $p_1$ is highly non-trivial. A proof of this fact was given in \cite{Eynard:2007}.

\begin{prop}
Let $\tilde{W}^g_n(\mathbf{p})$ be the correlation functions constructed from the local recursion, Definition \ref{d:recursion}, for an arbitrary spectral curve $\tilde{C}$. Then they are fully symmetric under permutations of the arguments.
\end{prop}

\begin{proof}
This is clear from Theorem \ref{t:main}. Indeed, we can obtain the correlation functions $\tilde{W}^g_n(\mathbf{p})$ by taking a succession of limits of correlation functions $W^g_n(\mathbf{p})$ obtained from the original recursion for a curve with only simple ramification points. Therefore, if the $W^g_n(\mathbf{p})$ are symmetric, as proved in \cite{Eynard:2007}, then the $\tilde{W}^g_n(\mathbf{p})$ are also symmetric.
\end{proof}

\subsection{Poles of the correlation functions}

\begin{prop}
Let $\tilde{W}^g_n(\mathbf{p})$ be the correlation functions constructed from the local recursion, Definition \ref{d:recursion}, for an arbitrary spectral curve $\tilde{C}$. Then the  $\tilde{W}^g_n(\mathbf{p})$ only have poles along $R$ in each variable (where $R$ is the set of ramification points). Moreover, if $a_i \in R$ has multiplicity $r_i$, then the order of the pole of $\tilde{W}^g_n(\mathbf{p})$ at $a_i$ in each variable is at most $(2g-2+n) r_i + 2g$.
\end{prop}

\begin{rem}
Note that for simple ramification points, $r_i=2$ and the poles are of order at most $(2g-2+n)2+2g = 6g-4+2n$, which was proved in \cite{Eynard:2007}.
\end{rem}

\begin{proof}
This follows by induction. First, it is clear from Definition \ref{d:recursion} that the only poles of the correlation functions $\tilde{W}^g_n(\mathbf{p})$ are along $R$.

To deduce the maximal order we proceed by induction. We will compute the maximal order of the pole in $p_0$; then by symmetry, the order must be the same in all variables.

Let us prove by induction that the order is at most $(2g-2+n) r_i+2g$. The initial conditions of the induction are $\tilde{W}^0_3$ and $\tilde{W}^1_1$. 

For $\tilde{W}^0_3$, the only pole in the integrand at $q=a_i$ comes from the pole of the recursion kernel $K_2$, which has order $r_i$. Hence $\tilde{W}^0_3$ can have a pole at $p_0=a_i$ of order at most $r_i = (2 (0) - 2 + 3) r_i + 2(0)$.

For $\tilde{W}^1_1$, the kernel provides a pole of order $r_i$ at $q=a_i$, and it is multiplied by a  term of the form $W^0_2(q, \theta_i^k(q))$ for some $k$, which provides an additional pole of order $2$. Hence $\tilde{W}^1_1$ can have a pole at $p_0=a_i$ of order at most $r_i+2 = (2(1)-2+1) r_i + 2(1)$.

Now consider $\tilde{W}^g_{n+1}$, and assume that all correlation functions $\tilde{W}^{g'}_{n'+1}$ with $2g'-2+n' < 2g-2+n$ have poles at $a_i$ of order at most $(2 g'-2+(n'+1))r_i + 2g'$. Consider one of the term in the recursion, corresponding to a subset $\mathbf{d}'_i(q)$ of $\mathbf{d}_i(q) = \{\theta_i^1(q), \ldots, \theta_i^{r_i-1}(q) \}.$ Let $k-1$ be the length of the subset $\mathbf{d}'_i(q)$; hence $1 \leq k-1 \leq r_i-1$.  Then the recursion kernel has a pole at $q=a_i$ of order $(k-1) r_i$. The kernel is multiplied by $\mathcal{W}^g_{k,n}$. Each term in $\mathcal{W}^g_{k,n}$ comes from a partition $\mu$ of $\{q, \mathbf{d}'_i(q) \}$. Then each such term has a pole at $q=a_i$ of order at most $r_i (2 (g+\ell(\mu)-k) -2 k + n + k) + 2 (g + \ell(\mu)-k)$. But $\ell(\mu) \leq k$, so the pole of $\mathcal{W}^g_{k,n}$ at $q=a_i$ is of order at most $r_i (2g-k+n) + 2 g$. Therefore, $\tilde{W}^g_{n+1}$ has a pole at $p_0=a_i$ of order at most $r_i (2g-1+n) + 2 g$. This is precisely the formula in the Proposition for $\tilde{W}^g_{n+1}$ (in terms of $n' = n+1$, it becomes $r_i (2g-2+n') + 2g$).

\end{proof}

\subsection{Dilaton equation}

Another important property of the correlation functions $W^g_n(\mathbf{p})$ that was proved in \cite{Eynard:2007} is that they satisfy the \emph{dilaton equation}:
\begin{equation}
W^g_{n}(\mathbf{p}) = \frac{1}{2g-2+n} \sum_{i=1}^m \underset{q=a_i}{\rm Res} \Phi(q) W^g_{n+1}(q,\mathbf{p}),
\end{equation}
where
\begin{equation}
\Phi(q) = \int^q_o y(t) \mathrm{d} x(t)
\end{equation}
with $o$ an arbitrary base point.

\begin{prop}
Let $\tilde{W}^g_n(\mathbf{p})$ be the correlation functions constructed from the local recursion, Definition \ref{d:recursion}, for an arbitrary spectral curve $\tilde{C}$. Then they satisfy the dilaton equation:
\begin{equation}
\tilde{W}^g_{n}(\mathbf{p}) = \frac{1}{2g-2+n} \sum_{i=1}^m \underset{q=a_i}{\rm Res} \tilde{\Phi}(q) \tilde{W}^g_{n+1}(q,\mathbf{p}).
\end{equation}
\end{prop}

\begin{proof}
Again, this follows from Theorem \ref{t:main}. We obtain the $\tilde{W}^g_n$ through successive limits of colliding ramification points from the $W^g_n$ for a curve with only simple ramification. Start with the dilaton equation satisfied by the $W^g_n$, as proved in \cite{Eynard:2007}:
\begin{equation}
W^g_{n}(\mathbf{p}) = \frac{1}{2g-2+n} \sum_{i=1}^k \underset{q=b_i}{\rm Res} \Phi(q) W^g_{n+1}(q,\mathbf{p}),
\end{equation}
where the $b_i$ are the ramification points of the starting curve with only simple ramification. The integrand is the same at each $b_i$, so the sum over residues is in fact a contour integral, where the contour is around all ramification points $b_i$ but no other poles of $\Phi$. We can then take the limit of colliding ramification points inside the contour integral; then $\Phi$ becomes the $\tilde{\Phi}$ of the limiting curve, the correlation functions $W^g_{n+1}$ become the $\tilde{W}^g_{n+1}$ of the limiting curve, and the contour integral becomes a sum over residues at the ramification points $a_i$ of the limiting curve. Therefore we obtain the dilaton equation for the $\tilde{W}^g_n$.
\end{proof}

\subsection{A formula for $W^0_3$}

It was shown in \cite{Eynard:2007} that the simplest stable correlation function $W^0_3(p_1,p_2,p_3)$ satisfies a nice formula:
\begin{equation}\label{eq:formula}
W^0_3(p_1,p_2,p_3) = \sum_{i=1}^m \underset{q=a_i}{\rm Res} \frac{W^0_2(q,p_1)W^0_2(q,p_2)W^0_2(q,p_3)}{\mathrm{d} x(q) \mathrm{d} y(q)}.
\end{equation}

\begin{prop}
Let $\tilde{W}^0_3(p_1,p_2,p_3)$ be the correlation function constructed from the local recursion, Definition \ref{d:recursion}, for an arbitrary spectral curve $\tilde{C}$. Then it satisfies:
\begin{equation}\label{eq:w03}
\tilde{W}^0_3(p_1,p_2,p_3) = \sum_{i=1}^m \underset{q=a_i}{\rm Res} \frac{\tilde{W}^0_2(q,p_1)\tilde{W}^0_2(q,p_2)\tilde{W}^0_2(q,p_3)}{\mathrm{d} \tilde{x}(q) \mathrm{d} \tilde{y}(q)}.
\end{equation}

\begin{proof}
It follows once more from Theorem \ref{t:main}. We obtain $\tilde{W}^0_3(p_1,p_2,p_3)$ through a succession of limits of colliding ramification points from the $W^0_3(p_1,p_2,p_3)$ for a spectral curve $C$ with only simple ramification. It was proved in \cite{Eynard:2007} that
\begin{equation}
W^0_3(p_1,p_2,p_3) = \sum_{i=1}^k \underset{q=b_i}{\rm Res} \frac{W^0_2(q,p_1)W^0_2(q,p_2)W^0_2(q,p_3)}{\mathrm{d} x(q) \mathrm{d} y(q)},
\end{equation}
where again the $b_i$ are the simple ramification points of the spectral curve $C$. The integrand is the same at all ramification points $b_i$, so we rewrite the sum over residues as a contour integral around all ramification points but no other poles. We then take the limit of colliding ramification points inside; $\mathrm{d} x$ and $\mathrm{d} y$ become the $\mathrm{d} \tilde{x}$ and $\mathrm{d} \tilde{y}$ of the limiting curve, and $W^0_2$ becomes the $\tilde{W}^0_2$ of the limiting curve. The contour integral picks up residues at the remaining ramification points $a_i$ of $\tilde{C}$, and we get the formula \eqref{eq:w03}.
\end{proof}
\end{prop}

\begin{rem}
We can in fact prove this proposition directly from the generalized recursion; this is done in the Appendix. The direct proof is interesting in that it illustrates how non-trivial these properties are for the generalized recursion.
\end{rem}

\subsection{Some particular residues}

Other useful properties that were proved in \cite{Eynard:2007} are
\begin{equation}
\underset{q=a}{\rm Res} \left(x^k(q) W^g_{n+1}(q,\mathbf{p})\right) = 0, \qquad \text{for $k=0,1$ and all $a \in R$,}
\end{equation}
and
\begin{equation}
\sum_{i=1}^m \underset{q=a_i}{\rm Res} \left(x^k(q) y(q) W^g_{n+1}(q,\mathbf{p})\right) = \frac{1}{2} \sum_{j=1}^d \mathrm{d}_{\tau_j(p)} \left(\frac{x(\tau_j(p))^k W^g_{n}(\mathbf{p})}{\mathrm{d}x(\tau_j(p))} \right), \qquad \text{for $k=0,1$.}
\end{equation}
In particular,
\begin{equation}
\sum_{i=1}^m \underset{q=a_i}{\rm Res} \left(x^k(q) y(q) W^g_{1}(q)\right) =0, \qquad \text{for $k=0,1$.}
\end{equation}

\begin{prop}
Let $\tilde{W}^g_n(\mathbf{p})$ be the correlation functions constructed from the local recursion, Definition \ref{d:recursion}, for an arbitrary spectral curve $\tilde{C}$. Then they satisfy
\begin{equation}
\underset{q=a}{\rm Res} \left(\tilde{x}^k(q) \tilde{W}^g_{n+1}(q,\mathbf{p})\right) = 0, \qquad \text{for $k=0,1$ and all $a \in R$,}
\end{equation}
and
\begin{equation}
\sum_{i=1}^m \underset{q=a_i}{\rm Res} \left(\tilde{x}^k(q) \tilde{y}(q) \tilde{W}^g_{n+1}(q,\mathbf{p})\right) = \frac{1}{2} \sum_{j=1}^d \mathrm{d}_{\tau_j(p)} \left(\frac{\tilde{x}(\tau_j(p))^k \tilde{W}^g_{n}(\mathbf{p})}{\mathrm{d}\tilde{x}(\tau_j(p))} \right), \qquad \text{for $k=0,1$.}
\end{equation}
\end{prop}

\begin{proof}
This follows from the exact same argument as in the proof of the previous propositions.
\end{proof}

\subsection{Diagrammatic rules}

Finally, we note that just as for the original recursion of \cite{Eynard:2007}, the local recursion in Definition \ref{d:recursion} admits a graphical representation. The details of the diagrammatic representation follow from the analysis in section 3.3 of \cite{BHLMR} from the point of view of degenerations of Riemann surfaces. Basically, the main difference with the graphical representation of the original recursion of \cite{Eynard:2007} is that the trivalent vertex associated to the recursion kernel $K_2(p_0; q_1,q_2)$, with edges marked by the entries $p_0, q_1$ and $q_2$, should be replaced by $(k+1)$-valent vertices associated to the recursion kernels $K_k(p_0; q_1, \ldots, q_k)$ with edges marked by the entries $p_, q_1, \ldots, q_k$. The diagrammatic rules corresponding to the combinatorics in Definition \ref{d:recursion} then follow directly from the analysis of degenerations of Riemann surfaces in section 3.3 of \cite{BHLMR}.



\section{Conclusion}

In this paper we constructed a global version of the recursion of \cite{Eynard:2007,Eynard:2008} generalized in \cite{BHLMR}. Through this global version, we showed that the recursion proposed in \cite{BHLMR} for spectral curves with arbitrary ramification is in fact a limit of the original recursion of \cite{Eynard:2007,Eynard:2008} for colliding ramification points. This allowed us to show that the correlation functions constructed from the local recursion in \cite{BHLMR}, or the global recursion that we proposed here, satisfy the same properties as the original correlation functions of \cite{Eynard:2007,Eynard:2008}.

An interesting question arises from the point of view of matrix models. In this paper, we studied the local recursion in Definition \ref{d:recursion} directly, and showed that it is a limit of the original recursion of \cite{Eynard:2007}, without reference to matrix models. But the original proposal for the generalized recursion in \cite{BHLMR} was inspired by the work of Prats-Ferrer \cite{Prats:2010}, where a recursion based on a spectral curve with a double ramification point was found by solving the loop equations of a particular Cauchy two-matrix model. Then a natural question arises: do matrix models with solution of the loop equations given by Definition \ref{d:recursion} exist? And what do they look like? We believe that such matrix models indeed exist, but as far as we know they have not been studied yet. This is a very interesting question that deserves further investigation.

From the standpoint of the recursion itself, one could ask whether having a globally defined integrand may be useful for other things. One obvious consequence of having a global recursion is that we can now compute the contour integral over $\Gamma(q)$ in Definition \ref{d:recursionglobal} by evaluating residues \emph{outside the contour}, since we are working on a compact Riemann surface. Moreover, since the integrand is globally defined, it can be written as a rational function of the generators of the function field of $S$, in our case $x$ and $y$. Combining these two observations could lead to a very efficient way of doing calculations.

Moreover, the global recursion may provide a new approach to prove and clarify statements such as $x-y$ invariance of the free energies (see \cite{EOxy,BHLMR,BS:2011}). Recall that $x-y$ invariance of the free energies is not universal, as discussed in \cite{BHLMR,BS:2011}.\footnote{For instance, if we replace the functions $x$ and $y$ of Example \ref{ex:2} by $x(t)=t+1/t$, $y(t) = (t-b)^2$, for $b \neq 0$, then $x-y$ invariance of the free energies does not hold \cite{BHLMR}.} It would be interesting to clarify for what spectral curves are the free energies $x-y$ invariant, or whether the definition of the free energies can be modified to ensure $x-y$ invariance for general spectral curves. It seems that problems with $x-y$ invariance arise when the one-form  $y \mathrm{d} x$ has residues; if it is a residueless one-form on a compact Riemann surface, then $x-y$ invariance seems to hold. It is then natural to ask whether the global recursion can be used to clarify these statements about $x-y$ invariance. This deserves further investigation.

\appendix

\section{A direct proof of a formula for $W^0_3$}

In this appendix we provide a direct proof that the $W^0_3$ constructed from the local recursion, Definition \ref{d:recursion}, for a curve with arbitrary ramification satisfies the formula \eqref{eq:formula}. Let $C$ be a spectral curve with zeroes of $\mathrm{d} x$ given by $\{a_1, \ldots, a_m\}$, each with multiplicity $r_i$, $i=1,\ldots,m$.

From Definition \ref{d:recursion}, we have:
\begin{equation}
W^0_{3}(p_0, p_1,p_2) = \frac{1}{2 \pi i} \sum_{i=1}^m \oint_{\Gamma_i(q)} \left(\sum_{d=1}^{r_i-1}  K_{2}(p_0;q,\theta_i^d(q)) \mathcal{W}^0_{2,2}(q,\theta_i^d(q); p_1,p_2 ) \right),
\end{equation}
which can be rewritten as
\begin{gather}
W^0_{3}(p_0, p_1,p_2) =  \sum_{i=1}^m  \mathop{{\rm Res}}_{q=a_i}  \sum_{d=1}^{r_i-1}K_{2}(p_0;q,\theta_i^d(q))  \Big( W^0_2(q,p_1) W^0_2(\theta_i^d(q),p_2) \nonumber\\
+ W^0_2(q,p_2) W^0_2(\theta_i^d(q),p_1)\Big).
\label{eq:ww}
\end{gather}

Let us compute the residue using the local coordinate near $a_i$:
\be
z=(x(q)-a_i)^{1/r_{a_i}},
\ee
\emph{i.e.}
\be
x(q) = a_i+z^{r_{a_i}}, \qquad \mathrm{d} x(q) = r_{a_i}\,z^{r_{a_i}-1}\,\mathrm{d} z.
\ee
The local deck transformations on $U_i$ near $a_i$ are then:
\be
\theta_i^d(z) = \alpha^d\,z, \qquad \alpha={\rm e}^{\frac{2\pi i}{r_{a_i}}}.
\ee
We then Taylor expand near $q\to a$:
\be
W^0_2(p,q) = \sum_{k=0}^\infty z^k\,B_k(p)\,\mathrm{d} z,
\ee
\be
y(q) = \sum_{k=0}^\infty y_k\,z^k.
\ee
In particular we have:\footnote{Note that we are using here the symmetrized version of the recursion kernel, which is equivalent to the recursion kernel defined in Definition \ref{d:kernel}.}
\be
\int_{\theta_i^d(q)}^q W^0_2(p, \cdot) = \sum_{k=0}^\infty \frac{1-\alpha^{d(k+1)}}{k+1}\,z^{k+1}\,B_k(p),
\ee
and 
\begin{align}
(y(q)-y(\theta_i^d(q)))\,\mathrm{d} x(q) =& r_{a_i}\,\sum_{k=0}^\infty y_k (1-\alpha^{kd})\,  z^{k+r_{a_i}-1}\,dz  \\
=& z^{r_{a_i}}\,r_{a_i}\,(1-\alpha^d) y_1\,\, \left(1+\sum_{k=1}^\infty \frac{y_k}{y_1}\, \frac{1-\alpha^{kd}}{1-\alpha^d}\,  z^{k-1} \right)\,\mathrm{d} z.
\end{align}
This means that:
\begin{gather}
K_2(p_1;q,\theta_i^d(q))
= \frac{1}{z^{r_{a_i}}\mathrm{d} z}\frac{1}{2r_{a_i}\,y_1}\, \sum_{n=0}^\infty \sum_{k=0}^\infty \sum_{l_1,\dots,l_n\geq 1} \Bigg(\frac{B_k(p_1)}{k+1}\,\frac{1-\alpha^{(k+1)d}}{1-\alpha^d}\,\nonumber\\
\prod_{j=1}^n \frac{y_{l_j}}{y_1} \, \frac{1-\alpha^{l_j d}}{1-\alpha^d}\,\, z^{k+1-n+\sum_j l_j}\Bigg).
\end{gather}
And thus the residue at $q=a_i$ appearing in \eqref{eq:ww} is
\begin{gather}
\frac{1}{2r_{a_i}\,y_1}\,\sum_n \sum_{k_1,k_2,k_3} \sum_{l_1,\dots,l_n}\, \mathop{{\rm Res}}_{z=0}\, \frac{z^{k_1+k_2+k_3+1-n+\sum_j l_j}\,\mathrm{d} z}{z^{r_{a_i}}}
\,\,B_{k_1}(p_1)B_{k_2}(p_2)B_{k_3}(p_3) \prod_{j=1}^n \frac{y_{l_j}}{y_1}\nonumber \\
  \sum_{d=1}^{r_{a_i}-1} \frac{\alpha^{(k_3+1)d}+\alpha^{(k_2+1)d}}{(k_1+1)}\,\frac{1-\alpha^{(k_1+1)d}}{1-\alpha^d} \prod_{j=1}^n  \frac{1-\alpha^{l_j d}}{1-\alpha^d}.
\end{gather}
This residue is non-vanishing only if:
\be
k_1+k_2+k_3+\sum_j (l_j-1) = r_{a_i}-2,
\ee
which implies in particular that $0\leq k_i\leq r_{a_i}-2$ and $1\leq l_j\leq r_{a_i}-1$.

Let us now perform the sum over $d=1,\dots, r_{a_i}-1$.
We write:
\be
\frac{1-\alpha^{k d}}{1-\alpha^d} =\sum_{m=0}^{k-1} \alpha^{md},
\ee
and thus:
\begin{gather}
 \sum_{d=1}^{r_{a_i}-1} \frac{\alpha^{(k_3+1)d}+\alpha^{(k_2+1)d}}{(k_1+1)}\,\frac{1-\alpha^{(k_1+1)d}}{1-\alpha^d} \prod_{j=1}^n  \frac{1-\alpha^{l_j d}}{1-\alpha^d} \nonumber\\
=
\sum_{d=1}^{r_{a_i}-1} \frac{\alpha^{(k_3+1)d}+\alpha^{(k_2+1)d}}{(k_1+1)}\,\sum_{m=0}^{k_1} \sum_{m_j=0}^{l_j-1} \alpha^{d(m+\sum_j m_j)}.
\end{gather}
We have:
\be
1\leq k_3+1+m+\sum_j m_j\leq 1+k_2+k_1+\sum_j (l_j-1) = r_{a_i}-1-k_2\leq r_{a_i}-1.
\ee
This implies that
\be
\sum_{d=1}^{r_{a_i}-1} \alpha^{d(k_3+1+m+\sum_j m_j)} = -1,
\ee
and thus
\be
 \sum_{d=1}^{r_{a_i}-1} \frac{\alpha^{(k_3+1)d}+\alpha^{(k_2+1)d}}{(k_1+1)}\,\frac{1-\alpha^{k_1d}}{1-\alpha^d} \prod_{j=1}^n  \frac{1-\alpha^{l_j d}}{1-\alpha^d} 
 = \frac{2(k_1+1)}{k_1+1}\,\prod_{j=1}^n l_j =2 \prod_{j=1}^n l_j .
 \ee
Eventually, we obtain:
\begin{gather}
 \mathop{{\rm Res}}_{q=a_i}  \sum_{d=1}^{r_i-1}K_{2}(p_0;q,\theta_i^d(q))  \Big( W^0_2(q,p_1) W^0_2(\theta_i^d(q),p_2)+ W^0_2(q,p_2) W^0_2(\theta_i^d(q),p_1)\Big) \nonumber\\
= \frac{1}{r_{a_i}\,y_1}\,\sum_n \sum_{k_1,k_2,k_3} \sum_{l_1,\dots,l_n}\, \mathop{{\rm Res}}_{z=0}\, \frac{z^{k_1+k_2+k_3+1-n+\sum_j l_j}\,\mathrm{d} z}{z^{r_{a_i}}}
\,\,B_{k_1}(p_1)B_{k_2}(p_2)B_{k_3}(p_3) \prod_{j=1}^n \frac{l_j\,y_{l_j}}{y_1}.
\end{gather}
This last quantity is clearly symmetric in $p_1,p_2,p_3$, and it is equal to:
\be
\mathop{{\rm Res}}_{q=a_i}\, \frac{B(q,p_1)B(q,p_2)B(q,p_3)}{\mathrm{d} x(q) \mathrm{d} y(q)}.
\ee

\end{document}